\documentclass[a4paper,11pt,twoside]{scrartcl}

\usepackage{graphicx,epsfig}
\usepackage{ifthen}
\usepackage{axodraw}
\usepackage{amsmath}
\graphicspath{{./}}

\usepackage{fancyhdr} 
\usepackage{calc}

\fancypagestyle{plain}{%
\fancyhead{} 
}
\makeatletter   
\def\cleardoublepage{
\clearpage
\if@twoside 
  \ifodd\c@page
  \else \hbox{} 
  \thispagestyle{empty} 
  \newpage 
    \if@twocolumn
      \hbox{}
      \newpage
    \fi
  \fi
\fi
} 
\makeatother

\newcommand{\<}{\langle}
\renewcommand{\>}{\rangle}
\newcommand{\beq}{\begin{equation}}
\newcommand{\eeq}{\end{equation}}
\newcommand{\beqn}{\begin{eqnarray}}
\newcommand{\eeqn}{\end{eqnarray}}

\newcommand{\nn}{\nonumber}

\newcommand{\ov}{\overline}

\newcommand{\be}{\beta}
\newcommand{\ga}{\gamma}

\newcommand{\eps}{\epsilon}

\newcommand{\la}{\lambda}
\newcommand{\te}{\theta}

\newcommand{\mc}{\mathcal}

\newcommand{\eg}{{\em e.g.} }
\newcommand{\ie}{{\em i.e.} }
\newcommand{\id}{1 \hspace{1.15mm} \!\!\!\!1}
\newcommand{\D}{\Delta}
\newcommand{\heff}{\mc H_{\rm eff}^{\D F = 2}}

\def\simge{\mathrel{\rlap{\raise 0.511ex \hbox{$>$}}{\lower 0.511ex \hbox{$\sim$}}}}
\def\simle{\mathrel{\rlap{\raise 0.511ex \hbox{$<$}}{\lower 0.511ex \hbox{$\sim$}}}} 

\newcommand{\mpar}[1]{
\marginpar{
\ifthenelse{\isodd{\value{page}}}
{\flushleft {\sffamily {\bfseries {\boldmath #1 \unboldmath}}}}
{\flushright {\sffamily {\bfseries {\boldmath #1 \unboldmath}}}}}
}

\begin{document}

\thispagestyle{empty}
\phantom{xxx}
\vskip0.3truecm
\begin{flushright}
 TUM-HEP-665/07
\end{flushright}
\vskip0.7truecm

\begin{center}
\LARGE{\boldmath \bfseries \sffamily 
The MFV limit of the MSSM for low $\tan \be$:\\
meson mixings revisited}

\vspace{1.1cm}

\noindent {\bfseries \sffamily \Large Wolfgang~Altmannshofer, Andrzej~J.~Buras \\ and Diego~Guadagnoli}

\vspace{1.1cm}

\normalsize

\noindent {\sl Physik Department, Technische Universit\"at M\"unchen,\\
James-Franck-Strasse, D-85748 Garching, Germany}

\vspace{0.5cm}

\end{center}

\vspace{0.4cm}

\begin{center}
{\bfseries \sffamily Abstract}
\end{center}
\begin{abstract}
\noindent We apply the effective field theory definition of Minimal Flavour Violation (MFV) 
to the MSSM. We explicitly show how, by this definition, the new sources of flavour and CP
violation present in the MSSM become functions of the SM Yukawa couplings, and cannot be simply set 
to zero, as is common wisdom in phenomenological MSSM studies that assume MFV.

\noindent We apply our approach to the MSSM $\Delta B = 2$ Hamiltonian at low $\tan \be$. 
The limit of MFV amounts to a striking increase in the predictivity of the model. In particular, 
SUSY corrections to meson-antimeson mass differences $\D M_{d,s}$ are always found to be 
positive with respect to the SM prediction. This feature is due to an interesting interplay
between chargino and gluino box diagrams (the dominant contributions) in the different mass
regimes one can consider.

\noindent Finally, we point out that, due to the presence of gluinos, the MFV MSSM does not 
belong -- even at low $\tan \be$ -- to the class of models with the so-called
`constrained' MFV (CMFV), in which only the SM operator $(V-A)\otimes(V-A)$ contributes to 
$\D M_{d,s}$. Consequently, for the MSSM and in the general case of MFV, one should not
use the Universal Unitarity Triangle (UUT), relevant for CMFV models, but a MFV-UT
constructed from $\be_{\psi K_S}$ and $|V_{ub}|$ or $\ga$ from tree-level decays. In
particular, with the measured value of $\be_{\psi K_S}$, MFV implies a testable correlation 
between $|V_{ub}|$ and $\ga$. With the present high value of $|V_{ub}|$, MFV favours $\ga
> 80^\circ$.
\end{abstract}

\newpage
\noindent \line(1,0){402.52}
\vspace{-0.3cm}
\tableofcontents
\noindent \line(1,0){402.52}

\section{Introduction} \label{sec:intro}

The analysis of accelerator data over the last two decades established the success 
of the Standard Model (SM) {\em pattern} of flavour and CP violation. If the New Physics 
(NP) introduced to stabilize the electroweak (EW) breaking scale had a sensibly different 
pattern for such violations, it is natural to expect that it would have been already visible 
in the considerable amount of precise data on flavour changing neutral current (FCNC)
processes available today. The coming years will show whether this picture is altered by
new data, in particular CP-violation in the $B_s$-system and rare $K$-decays, where large
non-CKM sources of flavour and CP-violation are still possible within the most popular
extensions of the SM. 

On the theoretical side, the success of the CKM description of FCNC processes triggered 
the idea that the dynamics responsible for the peculiar form of the SM Yukawa couplings 
may be relevant only at energy scales much higher than the 
one typically introduced to stabilize the EW breaking. Such high-energy dynamics would then 
generate only the Yukawa couplings present in the SM and no additional flavour violating 
structures. From the low-energy point of view, the SM Yukawa couplings are then the only `building
blocks' regulating the amount of FCNC and CP violating processes, and their form is then 
raised as a ``symmetry requirement'' for the flavour sector of any candidate 
extension of the SM at the EW scale \cite{MFV}.

The above mentioned idea is known in the literature as Minimal Flavour Violation. One
could say that this idea in the quark sector has become the more precise, the more 
experimental data tended toward it. A phenomenological definition of MFV, that uses (the
explicit occurrence of) the CKM matrix as the only source of flavour violation and
restricts the set of relevant operators in the low-energy effective Hamiltonian to the SM
ones, has been introduced in \cite{BurasMFV}. It implies a set of very special relations
\cite{BurasMFV,BurasZakopane} among observables in the flavour sector, that have been
extensively tested in the recent years. In particular, the unexpected agreement of the 
so-called Universal Unitarity Triangle \cite{BurasMFV} with the available data \cite{UTfit}
has brought MFV to the fore, raising the question how to implement it in NP models, whose 
flavour sector is {\em a priori unrelated} to the SM one.

Already at this stage, we would like to emphasize that, while pragmatic and
phenomenologically useful, the definition of MFV introduced in \cite{BurasMFV}, to be
called `constrained MFV' (CMFV) \cite{BBGT} in what follows, is not as general as the one in \cite{MFV}, 
and the difference between the two approaches will emerge from our discussion. For this
reason, in the present paper we will use the general definition of \cite{MFV} and only at
the end we will investigate under which assumptions the limit of CMFV can be reached in
the specific framework of the MSSM.

Stated loosely, the basic idea to provide a model independent definition of MFV is
as follows. The only low-energy remnant structures responsible for flavour violation are
the SM Yukawa couplings, the single ones `required' at present by experiments. Then, the flavour sector 
of every extension of the SM at the EW scale should be minimal flavour violating 
if its flavour violating `building blocks' are exclusively the SM Yukawa couplings. This idea has
been formulated rigorously by the authors of \cite{MFV}.

On the phenomenological side, the idea of MFV has very often been advocated to better 
constrain models, whose predictivity is spoiled by the large number of parameters, as is 
notably the case for the Minimal Supersymmetric Standard Model (MSSM).
Focusing on the latter, many studies do already exist in the literature, where the MFV
paradigm is explicitly advocated. However, many (most, actually) of such studies 
appeared before the ``effective theory'' definition by \cite{MFV} and use assumptions often 
not complying with such definition. We stress that the latter is the only one that
can be unambiguously applied in extensions of the SM, since MFV is defined through the
formal transformation properties of the SM Yukawa couplings and can subsequently be applied to any 
{\em new} source of flavour violation.

\bigskip

The aim of this paper is twofold. First, we reconsider the quark flavour sector of the 
MSSM and carefully discuss its MFV limit. We explicitly show how the new sources 
of flavour and CP violation present in the MSSM become in this limit functions of the SM 
Yukawa couplings, and cannot be simply dropped in MFV, as often assumed in the literature. A simplistic
but intuitive picture is that the off-diagonal entries in the soft terms -- the genuinely 
new sources of flavour violation in the MSSM -- are {\em not} zero in MFV, but instead `CKM-like'.

\noindent We then apply our approach to the specific case of the $\D B = 2$ Hamiltonian in the MSSM
at low $\tan \be$. Such Hamiltonian, responsible for $B_{s,d}$ meson mixings, provides a concrete
and phenomenologically interesting benchmark for the approach itself. Explicit implementation 
of MFV in the MSSM leads to a striking improvement of its predictivity. This is obvious if 
one thinks that the sector introducing the largest number of new parameters is notably the
{\em soft} sector. The latter is now entirely constrained to be proportional to appropriate 
combinations of the SM Yukawa couplings, so that the main unknowns turn out to be the (real) 
proportionality factors (`MFV parameters'), amounting to 12 independent dimensionless parameters. 
Furthermore, one has to fix some real mass scale parameters (`SUSY scales'): 
the $\mu$ parameter, a squark mass scale $\ov m$ and trilinear coupling $A$, gaugino 
masses $M_1, M_2$ and $M_{\tilde g}$ and the two Higgs soft terms $m_{H_u}$ and $m_{H_d}$.
Hence, the mass scales relevant to the $\D B = 2$ case are in total 8, but we will 
see that only a subset of them affects non-trivially the calculations. Finally, also 
$\tan \be$ is of course a parameter, but we set it to reference (small) values, whose 
choice does not affect our main findings.

\noindent Concerning the new sources of flavour violation specific to the $\D B = 2$ case, 
we note that contributions from boxes featuring gluinos and neutralinos, usually assumed not 
to enter MFV calculations \cite{gabrielli-giudice,BGGJS,BCRS-NP,BCRS-PL,BCRSbig}, do actually contribute. 
The flavour violating structures in their couplings are proportional, as mentioned above, 
to appropriate combinations of the SM Yukawa couplings, the combinations being fixed by the 
very definition of MFV \cite{MFV}.

The second aim of our paper is a detailed numerical study of the $\D B = 2$ Hamiltonian in
the MFV MSSM at low $\tan \be$. In this study, mass scale parameters can be fixed to
reference values covering all the physically interesting mass scenarios that low-scale
SUSY could have. Concretely, we fixed $\mu$ to a small (200 GeV) or intermediate (500
GeV) or large value (1000 GeV), covering both signs. For each choice of $\mu$ we then chose
the squark mass scale to four benchmark values in the range 100 $\div$ 1000 GeV and so forth
for the other parameters. We considered a total of 48 scenarios. Then, the 12 MFV parameters 
which govern proportionality to the Yukawa matrices are left free to float within
reasonable intervals. 

\noindent Now, for every mass scenario considered, a random scan of the MFV parameters 
allows to generate a range of predictions for the SUSY corrections to the SM 
meson-antimeson mass differences $\D M_{s,d}$. The predicted corrections display a number of remarkable 
features
\begin{itemize}
\item[{\bf i)}] For each of the mass scenarios considered, corrections turn out to be
always {\em positive} and to float within a relatively narrow range of values when varying MFV
parameters. Specifically, in the case of $B_s - \ov B_s$ mixing, corrections are in the
range $\D M_s^{\rm SUSY} \cong +(0 \div 2)~{\rm ps}^{-1}$. Given the still large error
associated with the computation of the matrix elements entering the $\D M_s$
determination, these corrections, at present, are however not large enough to distinguish
the SM from the MFV MSSM at low $\tan \beta$.
\item[{\bf ii)}] The positiveness of the sum of the SUSY contributions turns out to be
caused by the interplay between the two dominant of them, namely chargino and gluino
boxes. This interplay is mainly dictated by the relative importance of the $\mu$ parameter
with respect to the other SUSY scales. This can intuitively be understood by observing
that, if $\mu$ is small, it governs the chargino lightest mass eigenvalue, whereas large
values of $\mu$ increase the importance of scalar operator contributions coming from 
gluino boxes.
\item[{\bf iii)}] By analyzing the single box contributions, we identify four main
scenarios for the interplay between chargino and gluino contributions. Such scenarios are 
ruled basically by the magnitude of $\mu$ and by that of the squark mass scale
$\ov m$. Variation of the other SUSY scales plays only a marginal role in the qualitative
picture that emerges.
\item[{\bf iv)}] Since we restrict our analysis to low $\tan \be$, a naive expectation would 
be that most of the contributions be proportional to the SM left-left current operator, since
the down-quark Yukawa matrix should be negligible. We find departures from this picture, arising 
when $\mu$ is not small in magnitude, and due to gluino contributions. 
Responsible for these departures are, in particular, the LR and RR submatrices of the down-squark 
mass matrix. If the down Yukawa is set to zero, these submatrices are respectively zero or 
proportional to the identity matrix. On the other  hand, when the down Yukawa is kept, they 
give rise to the bulk of contributions from operators other than the SM one.
\end{itemize}

\medskip

The rest of the paper is organized as follows. In Section \ref{sec:MFV} we recall the
effective theory definition of MFV, in the formalism of the MSSM. Then, in Section
\ref{sec:MSSM-MFV}, we apply such definition to the flavour sector of the MSSM, in
particular to the soft SUSY breaking terms, by discussing their MFV relations to the SM
Yukawa couplings. In Section \ref{sec:MSSM-DF2} we then focus on the $\D B = 2$ Hamiltonian in the
MSSM at low $\tan \be$. We collect here all the basic formulae and discuss the steps needed
to evaluate their MFV limit in the light of the procedure described in the previous
sections. Section \ref{sec:MC} presents our numerical strategy to explore the MFV MSSM 
predictions for $\D M_{s,d}$ and a detailed accout of our main findings, 
in particular the features outlined in the above points {\bf i)} to {\bf iv)}. In Section
\ref{sec:discussion}, we then elaborate on our findings, by describing additional numerical 
studies performed to clarify the issues emerged, like the role of $\tan \be$. Section
\ref{sec:limits-MFV} is devoted to various considerations on the topic of MFV, triggered
by what we learned from the study carried out in the previous sections. One of such
reflections concerns the definition of the Universal Unitarity Triangle, which turns out
not to be a construction always valid in MFV. To this point we devote Section
\ref{sec:MFV-UUT}. Finally, Section \ref{sec:conclusions} presents our conclusions and outlook.
In the Appendix we collect the complete list of Wilson coefficients for the $\D B = 2$ 
Hamiltonian in the MSSM at low $\tan \be$.

\section{Minimal Flavour Violation: effective theory definition} \label{sec:MFV}

In a top-down approach, the question how to build up concretely the MFV hypothesis 
in a given NP framework translates into how to {\em define} MFV in the presence of new flavour
violating interactions, {\em a priori unrelated} to the SM ones. To understand this point,
it is first useful to remind the structure of flavour breaking in the SM. The SM flavour 
symmetry group and its breaking have been first elucidated 
in \cite{chivukula-georgi-MFV,hall-randall-MFV}.
Responsible for such breaking are the SM Yukawa couplings, and their transformation properties 
under the flavour group can be identified by requiring (formal) invariance of the Yukawa 
interactions. Flavour violation is recovered as the spurion Yukawa ``fields'' assume their 
background values. 
\noindent MFV then demands the Yukawa background values to be the {\em only} structures 
generating the observed flavour (and CP) violation. This definition, which has the
advantage to hold model-independently, has been introduced by D'Ambrosio {\em et al.}
\cite{MFV}.

Following this approach, in the context of a given NP model, every new flavour violating 
``coupling'' can be classified according to its transformation properties under the SM 
flavour group, and -- if MFV holds -- rewritten in terms of combinations of the SM 
Yukawa couplings transforming in the same way.

Such procedure has been detailed in \cite{MFV}. We now restate it briefly for the MSSM with 
$R$-parity, which is our case of interest. The relevant quantity is the superpotential $W$, 
which reads
\beqn
W = \eps_{ij} \left( Y^{IJ}_u H_u^i Q^{Ij} U^J 
+ Y^{IJ}_d H_d^i Q^{Ij} D^J + Y^{IJ}_e H_d^i L^{Ij} E^J 
+ \mu H_u^i H_d^j \right)~,
\label{W}
\eeqn
with the matter superfields $Q$, $U$, $D$, $L$, $E$ (containing SM fermions) and
$H_{u,d}$ (containing the Higgs doublets). Here $I,J$ and $i,j$ denote flavour and $SU(2)_L$
indices, respectively. The notation and conventions comply entirely with \cite{Rosiek}. 

Out of the largest possible group $G_F$ of field redefinitions that commutes with the gauge
group \cite{chivukula-georgi-MFV},
\beqn
G_F = \left[ SU(3) \otimes U(1) \right]^5 \equiv 
\underset{F=Q,U,D,L,E}{\bigotimes} \left[ SU(3) \otimes U(1) \right]_F~,
\label{GF}
\eeqn
the Yukawa interactions in the superpotential (\ref{W}) break the flavour group 
$[SU(3)]^5 \otimes U(1)_{E}$ \cite{peccei-quinn,MFV}.
The flavour symmetry can formally be recovered in eq. (\ref{W}) by treating $Y_{u,d,e}$ as
spurions and requiring them to have indices transforming under $[SU(3)]^5$ as follows
\beqn
\left[ Y_u \right]_{{\ov 3_Q}{3_U}}~, ~~~~\left[ Y_d \right]_{{\ov 3_Q}{3_D}}~, 
~~~~\left[ Y_e \right]_{{\ov 3_L}{3_E}}~,
\label{Yindices}
\eeqn
with the subscript $Q,U,D,L,E$ referring to an index that transforms as the corresponding
representation under $SU(3)_{Q,U,D,L,E}$, respectively, and as a singlet under 
all the other group factors. Note that the superfields $U,D,E$ are left-handed but must
describe right-handed particles. As a consequence their component fields are defined with
a charge conjugation operation and they transform as $\ov 3$ representations under 
$SU(3)_{U,D,E}$, respectively.

Using the $[SU(3)]^5$ symmetry, the fermion superfields can be suitably shifted to
have Yukawa couplings in the form
\beqn
Y_u = K^T \hat Y_u~, ~~~~ Y_d = \hat Y_d~, ~~~~ Y_e = \hat Y_e~,
\label{Ydiag}
\eeqn
with the $\hat Y$ diagonal matrices and $K$ the CKM matrix. This form is not the usual 
one, since quark mass matrices are not simultaneously diagonal. However, it is very 
useful when ranking different flavour changing effects, since the top Yukawa (the 
dominant one) displays explicit proportionality to the CKM matrix.

For low $\tan \be \equiv \< H_u \> / \< H_d \>$, all FCNC effects are dominantly 
described by one single off-diagonal structure \cite{MFV}
\beqn
(\la_{\rm FC})_{ij} \equiv
\left \{
\begin{array}{cl}
(Y_u Y_u^\dagger)_{ij} \approx \la_t^2 K_{3 i} K^*_{3 j}~, & i \neq j~, \\
[0.2cm]
0~, & i = j~,
\end{array}
\right.
\label{laFC}
\eeqn
with $\la_t = (\hat{Y}_u)_{33}$.
Note in fact that higher powers of $Y_u Y_u^\dagger$ can be rewritten in terms of 
$Y_u Y_u^\dagger$ times an appropriate power of $\la_t^2$.
Subleading effects on the r.h.s. of eq. (\ref{laFC}) are suppressed by powers of 
$m_c/m_t$. 

The main observation \cite{MFV} is now that, if MFV holds, soft SUSY-breaking terms are
related to the SM Yukawa couplings (\ref{Ydiag}) and the explicit relations can be constructed 
by just using the formal transformation properties of (\ref{laFC}) under the flavour group. 
Such derivation will be presented in section \ref{sec:MSSM-MFV}. In section \ref{sec:MSSM-DF2} 
we will subsequently use the calculated SUSY parameters to evaluate the MSSM contributions 
to the $\D B = 2$ effective Hamiltonian, to which we will restrict the rest of the analysis. 
The latter will allow us to quantitatively assess the features of the SUSY contributions 
in the MFV limit in a benchmark case like meson oscillations.

In this respect, we anticipate that even if $(\la_{\rm FC})_{ij}$ in eq. (\ref{laFC}) provides 
the dominant FC mechanism, a detailed study shows that effects proportional to the 
down-quark Yukawa cannot actually be neglected. The latter does not provide by itself an
additional FC mechanism -- as one can see from eq. (\ref{Ydiag}) -- but still its effects can correct
the magnitude of those provided by $(\la_{\rm FC})_{ij}$. This can be understood by simply looking at the
LR entries of the down-squark mass matrix (see eq. (\ref{Md}) below). The latter are 
proportional to $\mu Y_d$, and if $\mu$ is not small, the corresponding terms cannot be
dropped even if $Y_d \ll Y_u$.

Focusing again on the $\D B = 2$ Hamiltonian, this mechanism also regulates the relative 
importance of contributions to operators beyond the SM left-left vector operator $\mc Q_1$. 
This can also be naively understood from the down-squark mass matrix, once its MFV limit is
performed. In this limit, in fact, the LR and RR sectors are zero or respectively
proportional to the identity matrix, if $Y_d \to 0$. As a consequence one would expect the 
SUSY contributions to the $\D B = 2$ Hamiltonian be in the Wilson coefficient of $\mc Q_1$ only. 
This picture is largely modified or completely spoiled when one includes $Y_d$, depending on 
the mass scenario chosen for the SUSY parameters. The main actor in this respect is again 
the $\mu$ parameter, for the reason outlined above.

\section{MFV relations of MSSM parameters to SM Yukawa couplings} \label{sec:MSSM-MFV}

We now discuss how the above picture concretely applies to the MSSM with $R$-parity and softly 
broken SUSY. In this model, the part of the Lagrangian responsible for flavour violation reads
\beqn
\mc L_{\rm MSSM}^{\rm f.v.} = [W]_{\te \te} + c.c. + \mc L_{\rm soft}^{\rm f.v.}~,
\label{LMSSMfv}
\eeqn
with $W$ the superpotential of eq. (\ref{W}) and $\mc L_{\rm soft}^{\rm f.v.}$ given as
\beqn
-\mc{L}^{\rm f.v.}_{\rm soft}= &&\Bigl[ 
(m^{IJ}_Q)^2 \left( (\tilde{u}^{I}_L)^* \tilde{u}^J_L
+ (\tilde{d}^I_L)^* \tilde{d}^J_L \right)
+ (m^{IJ}_U)^2 \tilde{u}^I_R (\tilde{u}^{J}_R)^* 
+ (m^{IJ}_D)^2 \tilde{d}^I_R (\tilde{d}^{J}_R)^* \nn \\
&&+(m^{IJ}_L)^2 \left( (\tilde{\nu}^I_L)^* \tilde{\nu}^J_L
+ (\tilde{e}^I_L)^* \tilde{e}^J_L \right) 
+ (m^{IJ}_E)^2 \tilde{e}^I_R (\tilde{e}^{J}_R)^* \nn \\
&&+(m_{H_u})^2 \left( (h^1_u)^* h^1_u + (h^2_u)^* h^2_u \right) 
+ (m_{H_d})^2 \left( (h^1_d)^* h^1_d + (h^2_d)^* h^2_d \right) \Bigl] \nn \\
[0.2cm]
&&- \Bigl[ \eps_{ij} \Bigl( A^{IJ}_u h_u^i \tilde{q}_L^{Ij} (\tilde{u}^J_R)^*
+ A^{IJ}_d h_d^i \tilde{q}_L^{Ij} (\tilde{d}^J_R)^* \nn \\
&&+ A^{IJ}_e h_d^i \tilde{\ell}_L^{Ij} (\tilde{e}^J_R)^*
 + B_h h_u^i h_d^j \Bigl) ~+~ c.c. \Bigl] ~,
\label{LMSSMsoft}
\eeqn
\ie the usual soft Lagrangian with omitted gaugino mass terms.

The MSSM Lagrangian, with the above superpotential and soft pieces, gives rise to the
following $6 \times 6$ squark mass matrices
\beqn
M_{\tilde u}^2 =
\left(
\begin{array}{cc}
\frac{v_2^2}{2} Y_u Y_u^\dagger + (m_Q^2)^T - \frac{\cos 2 \be}{6}( M_Z^2 - 4 M_W^2 ) \id & 
- \mu^* \frac{v_1}{\sqrt 2} Y_u - \frac{v_1}{\sqrt 2} \tan \be A_u\\
[0.2cm]
- \mu \frac{v_1}{\sqrt 2} Y_u^\dagger - \frac{v_1}{\sqrt 2} \tan \be A_u^\dagger& 
\frac{v_2^2}{2} Y_u^\dagger Y_u + m_U^2 + \frac{2}{3} \cos 2 \be M_Z^2 s_{\rm w}^2 \id
\end{array}
\right)~,\nn \\
\label{Mu}
\eeqn
\beqn
M_{\tilde d}^2 =
\left(
\begin{array}{cc}
\frac{v_1^2}{2} Y_d Y_d^\dagger + (m_Q^2)^T - \frac{\cos 2 \be}{6}( M_Z^2 + 2 M_W^2 ) \id & 
\mu^* \frac{v_1}{\sqrt 2} \tan \be Y_d + \frac{v_1}{\sqrt 2} A_d\\
[0.2cm]
\mu \frac{v_1}{\sqrt 2} \tan \be Y_d^\dagger + \frac{v_1}{\sqrt 2} A_d^\dagger& 
\frac{v_1^2}{2} Y_d^\dagger Y_d + m_D^2 - \frac{\cos 2 \be}{3} M_Z^2 s_{\rm w}^2 \id
\end{array}
\right)~.\nn \\
\label{Md}
\eeqn

If MFV holds, the new sources of flavour violation present in the soft terms must be related 
to SM Yukawa couplings. To this end, they can again be formally treated as spurion fields, with indices 
transforming under the flavour group as follows:
\beqn
[m_Q^2]_{3_Q \ov 3_Q}~, ~~ [m_U^2]_{\ov 3_U 3_U}~, ~~[m_D^2]_{\ov 3_D 3_D}~,
 ~~[A_u]_{\ov 3_Q 3_U}~, ~~[A_d]_{\ov 3_Q 3_D}~.
\eeqn
Recalling Yukawa transformations (\ref{Yindices}), one can then write the following MFV relations 
\cite{MFV}\footnote{Note that $b_3$ and $b_4$ must be equal, due to the hermiticity of
$m_Q^2$.\label{b3b4}}
\beqn
[m_Q^2]^T &=& \ov m^2 \left( a_1 \id + b_1 Y_u Y_u^\dagger + b_2 Y_d Y_d^\dagger 
+ b_3 Y_d Y_d^\dagger Y_u Y_u^\dagger + b_4 Y_u Y_u^\dagger Y_d Y_d^\dagger \right)~, \nn\\
m_U^2 &=& \ov m^2 \left( a_2 \id + b_5 Y_u^\dagger Y_u \right)~, \nn \\
m_D^2 &=& \ov m^2 \left( a_3 \id + b_6 Y_d^\dagger Y_d \right)~, \nn \\
A_u &=& A \left( a_4 Y_u + b_7 Y_d Y_d^\dagger Y_u \right)~, \nn \\
A_d &=& A \left( a_5 Y_d + b_8 Y_u Y_u^\dagger Y_d \right)~,
\label{softMFV}
\eeqn
where, in the first line, we have reported the transpose of $m_Q^2$, since it is the latter
to appear in the squark mass matrix in the usual conventions \cite{Rosiek}. The $a_i, b_i$ 
coefficients are real proportionality factors, whose allowed range of values will be
studied in section \ref{sec:MC} below. The overall mass scales $\ov m$ and $A$ fix
the order of magnitude of the respective soft terms, when the coefficient multiplying them
is of O(1). Expansions (\ref{softMFV}) are accurate to all orders in $\la_t$ and $\la_b
\equiv (\hat Y_d)_{33}$. Subleading effects are suppressed by powers of $m_c/m_t$
and/or $m_s/m_b$.

Starting from the mass matrices (\ref{Mu})-(\ref{Md}), and expressing the soft terms
according to expansions (\ref{softMFV}), it is then customary to perform a 
superfield redefinition leading to diagonal mass matrices for the quarks. The unitary 
matrices adopted are the same as in the SM: here one defines shifts
\beqn
&&u_L \to V_{Q_1} u_L~, ~~ d_L \to V_{Q_2} d_L~,\nn \\
&&u_R \to V_{U} u_R~, ~~~ d_R \to V_{D} d_R~,
\label{sCKM} 
\eeqn
and the fields on the r.h.s. form the CKM basis. In the MSSM, the same shifts are carried
out at the superfield level and lead to the so-called super-CKM basis. After performing such 
transformations, one gets diagonal Yukawa matrices $\hat Y_u, \hat Y_d$, and can use relations 
\beqn
\hat m_u = \frac{v_2}{\sqrt 2} \hat Y_u~,~~\hat m_d = -\frac{v_1}{\sqrt 2} \hat Y_d~,
\label{mqY}
\eeqn
in eqs. (\ref{Mu})-(\ref{Md}) to display explicit dependence on the quark mass matrices $\hat m_{u,d}$.

In the super-CKM basis, the matrices $m_{Q,U,D}^2$ and $A_{u,d}$
have still off-diagonal entries.\footnote{Such entries are responsible in general for 
genuinely supersymmetric flavour violation in the mass matrices (\ref{Mu})-(\ref{Md}). 
In our case, as we said, soft term are instead fixed by the MFV expansions (\ref{softMFV}).}
Then, in order to have the (hermitian) mass matrices in eqs. (\ref{Mu})-(\ref{Md}) in diagonal form,
one needs a second redefinition, performed on the up- and down-squark fields, respectively. 
Such redefinition leads from the basis $\tilde u, \tilde d$, to the mass eigenstate basis $U,D$, 
which in the conventions of \cite{Rosiek} reads
\beqn
\tilde u_i = (Z_U)_{ij} U_j~,~~
\tilde d_i = (Z_D^*)_{ij} D_j~,
\label{Zud}
\eeqn
with the index $i = 1,2,3$ for $\tilde u_L, \tilde d_L$ and $i = 4,5,6$ for 
$\tilde u_R, \tilde d_R$. 

Using transformation (\ref{Zud}), the down-squark mass term in the Lagrangian is
diagonalized according to
\beqn
\tilde d^T \, M_{\tilde d}^2 \, \tilde d^*
= D^T \, [ Z_D^\dagger M_{\tilde d}^2 Z_D ] \, D^*
= D^T \, \hat M_D^2 \, D^*~, ~~~ \hat M_D^2 = {\rm diag}\{M_{D_1}^2,\,..., M_{D_6}^2\}~,
\label{MdtoMD}
\eeqn
with $M_{\tilde d}^2$ given in eq. (\ref{Md}) and $\tilde d$ ($D$) a column vector built
out of the $\tilde d_i$ ($D_i$). An entirely analogous equation holds for the case of up-squarks.

\medskip

In practical calculations, flavour violation is driven either by non-diagonal squark
propagator matrices, when working in the $\tilde d, \tilde u$ basis, or by a $Z_{D,U}$ matrix 
appearing in vertices with a squark leg (gluino-quark-down squark, neutralino-quark-down
squark and chargino-quark-up squark, in our case).
In the former case, one usually adopts an expansion in off-diagonal ``mass insertions''
and stops to the first non-trivial order, in the well-know Mass Insertion Approximation
(MIA) \cite{MIA,GGMS}. The MIA provides a very useful tool to make flavour violation
mediated by soft SUSY breaking terms most transparent and manageable, since it
`linearizes' the mechanism of flavour violation, but it is an approximation.

In our MFV formulae, we will instead stick to the mass eigenstate basis, \ie to the exact
calculation. Among the $M_{\tilde d}^2$ entries in eq. (\ref{Md}), the soft terms will be
related through proportionality factors to the SM Yukawa couplings, according to the expansions
(\ref{softMFV}). Then the $M_{\tilde d}^2$ mass matrix turns out to depend only on $\mu$
(which must be real), on $\tan \be$, on the two squark scale factors $A, \ov m$ 
(see eq. (\ref{softMFV})) and on the proportionality factors. Upon rotations of the squark
states from the super-CKM basis to the mass eigenbasis, eqs. (\ref{Zud}), the pattern of
flavour violation is then transferred from the non-diagonality of the mass matrices, to
the off-diagonal entries of the matrices $Z_{D,U}$, entering quark-squark interactions
with gluinos and neutralinos.

Let us show with a simple example how flavour violation in the $Z_{D,U}$ becomes `CKM-like', 
after MFV expansions are imposed.
Let us consider the down-squark mass matrix of eq. (\ref{Md}), with soft SUSY parameters
given according to the MFV expansions in eq. (\ref{softMFV}). Adopting the approximation
$Y_d \to 0$, one can drop all the corresponding terms in such expansions. One can then
perform the super-CKM rotation on the squark fields to have the up Yukawa diagonal. The
down-squark mass matrix assumes, in this basis and under these assumptions, the following form
\beqn
M_{\tilde d}^2 =
\left(
\begin{array}{cc}
\ov m^2(a_1 \id + b_1 (K^\dagger \hat Y^2_u K)^T) 
- \frac{\cos 2 \be}{6}( M_Z^2 + 2 M_W^2 ) \id & 0\\
[0.2cm]
0 & \ov m^2 a_3 \id - \frac{\cos 2 \be}{3} M_Z^2 s_{\rm w}^2 \id
\end{array}
\right)~,\nn \\
\label{Mdexample}
\eeqn
whence the unitary transformation (\ref{Zud}) leading to the mass eigenbasis for the
down-squarks is obviously
\beqn
Z_{D} =
\left(
\begin{array}{cc}
K^T & 0 \\
0 & \id
\end{array}
\right)~.
\eeqn
As one can see, off-diagonal entries in $Z_D$ are not zero, but CKM-like (in this simple
case only in the LL sector). They will appear in the couplings of gluinos and neutralinos
with quarks and squarks.

When one includes in the diagonalization the effects of the down-Yukawa matrix, the
diagonalization becomes more involved. However, flavour violation in the $Z_{D,U}$ is
still encoded in their dependence on the SM Yukawa couplings. 

This observation allows us to comment on a conventional assumption present in most of the 
calculations performed in the MFV MSSM to date. This assumption amounts to dropping altogether 
flavour violating entries in the $Z_{D,U}$, due to the common wisdom that, if MFV holds, flavour violation 
can come only from couplings explicitly displaying proportionality to the CKM matrix.
The effective theory definition of MFV \cite{MFV} implies that the correct approach 
is instead to think the off-diagonal terms in the $Z_{D,U}$ as being not zero, but instead
dictated by the SM Yukawa couplings. In the simple example above, this dependence reconstructs 
in (the LL sector of) $Z_{D}$ directly the CKM matrix. 

The bottom line is that, in the MFV MSSM, one has to diagonalize squark matrices {\em after} 
imposing expansions (\ref{softMFV}), so that the diagonalization matrices $Z_{D,U}$ bear
dependence on such expansions and then use the $Z_{D,U}$ in all vertices where they are
required. This point has already been stated in \cite{MFV}.

\medskip

For the sake of completeness, we also report here the chargino, neutralino and charged
Higgs mass matrices, since these particles will enter our subsequent calculations. 
The notation used is again that of \cite{Rosiek}. Charginos are 
two Dirac fermions $\chi_{1,2}$ whose mass matrix reads
\beqn
\left(
\begin{array}{cc}
M_{\chi_1} & 0 \\
0 & M_{\chi_2}
\end{array}
\right) = 
Z_-^T
\left(
\begin{array}{cc}
M_2 & \frac{e v_2}{\sqrt 2 s_W}\\
\frac{e v_1}{\sqrt 2 s_W} & \mu
\end{array}
\right)
Z_+~,
\label{MCha}
\eeqn
with $Z_{\pm}$ unitary matrices, chosen from the requirement $0 < M_{\chi_1} < M_{\chi_2}$. 
Similarly, neutralinos are four Majorana fermions $\chi^0_{1,...,4}$, with mass matrix given by
\beqn
\left(
\begin{array}{ccc}
M_{\chi_1^0} & & 0 \\
 & \ddots &  \\
0 & & M_{\chi_4^0}
\end{array}
\right) = 
Z_N^T
\left(
\begin{array}{cccc}
M_1 & 0 & \frac{-e v_1}{\sqrt 2 c_W} & \frac{e v_2}{\sqrt 2 c_W}\\
0 & M_2 & \frac{e v_1}{\sqrt 2 s_W} &  \frac{-e v_2}{\sqrt 2 s_W}\\
\frac{-e v_1}{\sqrt 2 c_W} & \frac{e v_1}{\sqrt 2 s_W} & 0 & -\mu\\
\frac{e v_2}{\sqrt 2 c_W} & \frac{-e v_2}{\sqrt 2 s_W} & -\mu & 0
\end{array}
\right)
Z_N~,
\label{MNeu}
\eeqn
with $Z_N$ a unitary matrix, whose form is again specified after requiring positiveness 
and ordering for the eigenvalues.

\noindent Finally, one has two physical charged Higgs scalars $H_1^{\pm}$, with mass
\beqn
M_{H_1^\pm}^2 = M_W^2 + m_{H_u}^2 + m_{H_d}^2 + 2 |\mu|^2~,
\label{MH1}
\eeqn
where $m_{H_u}^2$ and $m_{H_d}^2$ are soft terms for the corresponding Higgs doublets,
given in eq. (\ref{LMSSMsoft}). Away from the unitary gauge, one must also include in the 
calculations the $H_2^{\pm}$ fields, which provide the longitudinal degrees of freedom 
for the $W$ bosons in the unitary gauge.

When assuming MFV, the gaugino masses $M_{1,2}$ are real.\footnote{$M_{\tilde g}$ can be
chosen as real without loss of generality \cite{Rosiek}.} In fact, if one allows
non-trivial phases in $M_{1,2}$, they are communicated to the diagonalization matrices
$Z_N$ and $Z_{\pm}$, which in turn enter Feynman rules for charginos and neutralinos. One
would then have new sources of CP violation, not allowed by the MFV hypothesis.
The same argument applies to the Higgs sector parameter $\mu$.

\section{\boldmath $\D B = 2$ in the MFV MSSM at low $\tan \be$} \label{sec:MSSM-DF2}

The MFV limit of the MSSM, as described in the previous sections, can now be applied 
to a concrete example, that of the $\D B = 2$ Hamiltonian, which is responsible for 
meson-antimeson oscillations. The latter has recently received renewed theoretical
interest, in view of the very precise measurement of $B_s - \ov B_s$ oscillations 
by the CDF collaboration \cite{CDF-DMs}.

The basic ingredient to describe meson-antimeson oscillations is the quantity 
$\mc M^{(M)}_{12}$ $\equiv$ $\< M | \heff | \ov M \>$, with $M = K$, $B_{d,s}$.
Within the MSSM, $\heff$ has the form
\beqn
\heff = \sum_{i=1}^5 C_i \mc Q_i + \sum_{i=1}^3 \tilde C_i \tilde {\mc Q}_i ~+{\rm H.c.}~,
\label{heff-MSSM}
\eeqn
with the $\mc Q_i$ given, in the case of $\ov B_s - B_s$ mixing, by
\beqn
\mc Q_1 & = & 
(\ov s^{i} \gamma_{\mu\,L}\, b^{i}) \, (\ov s^{j} \gamma^\mu_{L}\, b^{j})~,\nn \\
\mc Q_2 & = & 
(\ov s^{i} P_{L}\, b^{i})\, (\ov s^{j} P_{L}\, b^{j})~,\nn \\
\mc Q_3 & = & 
(\ov s^{i} P_{L}\, b^{j})\, (\ov s^{j} P_{L}\, b^{i})~,\nn\\
\mc Q_4 & = & 
(\ov s^{i} P_{L}\, b^{i})\, (\ov s^{j} P_{R}\, b^{j})~,\nn \\
\mc Q_5 & = & 
(\ov s^{i} P_{L}\, b^{j})\, (\ov s^{j} P_{R}\, b^{i})~.
\label{Qbasis}
\eeqn
The operators $\tilde{\mc Q}_{1,2,3}$ are obtained from $\mc Q_{1,2,3}$ 
by the replacement $L \to R$. The left- and right-handed projectors are defined 
as $P_{R,L}= (1\pm\gamma_5)/2$ and $\gamma^\mu_{R,L}=\gamma^\mu P_{R,L}$; 
$i,j$ are colour indices. In the case of $B_d$, one should replace $s \rightarrow d$
 in eq. (\ref{Qbasis}).

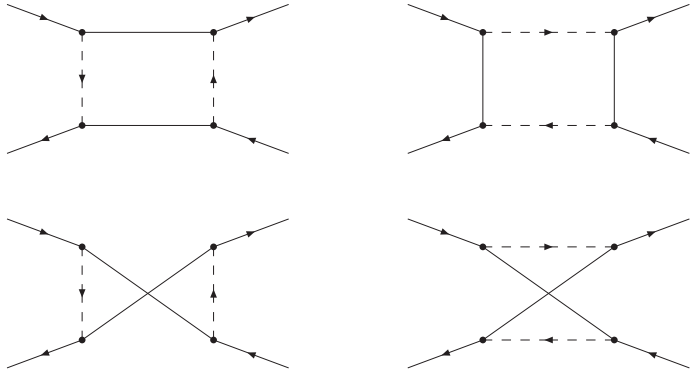
\begin{figure}
\vspace{-0.3cm}
\begin{center}
\SetScale{0.7}
\begin{picture}(150,80)(0,0)
\Vertex(40,15){1.8}
\Vertex(40,65){1.8}
\Vertex(110,15){1.8}
\Vertex(110,65){1.8}
\ArrowLine(40,15)(0,0)
\ArrowLine(0,80)(40,65)
\ArrowLine(150,0)(110,15)
\ArrowLine(110,65)(150,80)
\Line(40,65)(110,65)
\Line(110,15)(40,15)
\DashArrowLine(110,15)(110,65){5}
\DashArrowLine(40,65)(40,15){5}
\end{picture}
\begin{picture}(150,80)(0,0)
\Vertex(40,15){1.8}
\Vertex(40,65){1.8}
\Vertex(110,15){1.8}
\Vertex(110,65){1.8}
\ArrowLine(40,15)(0,0)
\ArrowLine(0,80)(40,65)
\ArrowLine(150,0)(110,15)
\ArrowLine(110,65)(150,80)
\DashArrowLine(40,65)(110,65){5}
\DashArrowLine(110,15)(40,15){5}
\Line(110,15)(110,65)
\Line(40,15)(40,65)
\end{picture}

\begin{picture}(150,80)(0,0)
\Vertex(40,15){1.8}
\Vertex(40,65){1.8}
\Vertex(110,15){1.8}
\Vertex(110,65){1.8}
\ArrowLine(40,15)(0,0)
\ArrowLine(0,80)(40,65)
\ArrowLine(150,0)(110,15)
\ArrowLine(110,65)(150,80)
\Line(40,65)(110,15)
\Line(110,65)(40,15)
\DashArrowLine(110,15)(110,65){5}
\DashArrowLine(40,65)(40,15){5}
\end{picture}
\begin{picture}(150,80)(0,0)
\Vertex(40,15){1.8}
\Vertex(40,65){1.8}
\Vertex(110,15){1.8}
\Vertex(110,65){1.8}
\ArrowLine(40,15)(0,0)
\ArrowLine(0,80)(40,65)
\ArrowLine(150,0)(110,15)
\ArrowLine(110,65)(150,80)
\DashArrowLine(40,65)(110,65){5}
\DashArrowLine(110,15)(40,15){5}
\Line(110,15)(40,65)
\Line(40,15)(110,65)
\end{picture}
\end{center}
\caption{\small\sl Feynman diagrams describing meson-antimeson oscillations 
in the MSSM. The crossed diagrams (second row) are needed only if the 
fermion in the loop is a Majorana particle. The notation for the various 
lines is the same as in \cite{Rosiek}.}
\label{fig:boxes}
\end{figure}

Each of the Wilson coefficients in eq. (\ref{heff-MSSM}) features, 
for low $\tan \beta$, the following contributions
\beqn
C_i = C_i^{\rm SM} + C_i^{H^+ H^+} + C_i^{\chi^+ \chi^+}  
+ C_i^{\tilde g \tilde g} + C_i^{\tilde g \chi^0} + C_i^{\chi^0 \chi^0} ~,
\label{Ci-MSSM}
\eeqn
where, for simplicity, we have omitted to specify the flavour indices of the 
external quarks, as in eq. (\ref{heff-MSSM}). In eq. (\ref{Ci-MSSM}), the first term on
the r.h.s. represents contributions from the SM boxes. The additional contributions, that
need to be considered within the MSSM, come respectively from boxes with 
charged Higgs-up quarks, chargino-up squarks, gluino-down squarks, mixed gluino- and 
neutralino-down squarks, and neutralino-down squarks. The possible Feynman diagrams involved 
in each case are represented in Fig.~\ref{fig:boxes}. For the SM, charged Higgs and chargino
cases, one has Dirac fermions propagating in the diagrams, so that only the boxes in the 
first row of Fig.~\ref{fig:boxes} must be considered. The other contributions involve Majorana
fermions in the loop, so that also crossed boxes (second row of Fig.~\ref{fig:boxes})
need to be calculated. 

The complete list of Wilson coefficients for the $\D B = 2$ effective Hamiltonian in 
the MSSM is reported explicitly in the Appendix.
Such coefficients are calculated in terms of loop functions, depending on the particle 
masses involved in the loops, and couplings, possibly featuring the rotation matrices 
$Z_{U,D}$ (squarks), $Z_{\pm}$ (charginos), $Z_N$ (neutralinos) introduced in the 
previous Section to define the respective mass eigenstates. To evaluate the Wilson
coefficients in MFV, the procedure to follow is now clear:
\begin{enumerate}
\item Expand the soft terms as in eq. (\ref{softMFV}) if they transform 
non-trivially under the flavour group, or take them as real if they are singlets;
\item Plug them into the mass matrices and diagonalize the latter to obtain the mass 
eigenvalues and the rotation matrices defining the eigenbases;
\item Use the obtained eigenvalues and rotation matrices in the general MSSM 
formulae for the Wilson coefficients.
\end{enumerate}

Now that we have all the ingredients of the calculation, we conclude by listing 
the number of parameters involved in the MFV limit of the $\D B = 2$ MSSM 
Hamiltonian for low $\tan \be$. The expansions of the soft terms in the squark 
mass matrices, eq. (\ref{softMFV}), 
involve 12 real proportionality factors, and 2 overall mass scales: a `generic' 
squark mass $\ov m$ and a `generic' trilinear mass term $A$. In addition one has 
to fix three real gaugino mass terms $M_1$, $M_2$ and $M_{\tilde g}$ and the 
real $\mu$ parameter. Finally, the soft Higgs sector adds 2 more mass 
scales, namely $m_{H_u}$ and $m_{H_d}$. Taking into account the requirement of correct
EW symmetry breaking, amounting to one constraint, one has a total of 12 + 7 
parameters.\footnote{We note that soft terms, expanded according to eq. (\ref{softMFV}), do actually
depend on the product between a mass scale and a MFV coefficient, so that the real
parametric dependence is on the product between the two. Considering this, the above
counting is somehow an overcounting.}
Among the latter, actually the dependence of the computed Hamiltonian on 
$m_{H_{u,d}}$ can be trivially `factored out' and not considered in MonteCarlo approaches.
In fact, $m_{H_{u,d}}$ enter only Higgs boxes, and the latter in turn depend exclusively on
$m_{H_{u,d}}$ and on $\mu$, through eq. (\ref{MH1}), which fixes the physical charged Higgs mass 
$M_{H_1^{\pm}}$. There is instead no dependence on any of the other SUSY scales and on the 
MFV parameters. In addition, considering that $m_{H_{u,d}}^2$ are demanded by the soft
Lagrangian (\ref{LMSSMsoft}) to be real, but not necessarily positive, it is clear that, for 
every choice of $\mu$, it is always possible to tune $m_{H_{u,d}}^2$ (compatibly with the
EW symmetry breaking constraint) so that $M_{H_1^{\pm}}$ assumes any desired value. As a 
consequence, one can trade the parametric dependence on $m_{H_{u,d}}^2$ for that on
$M_{H_1^{\pm}}$, which is then the only SUSY parameter in the Higgs contributions.

In the next section, we will discuss the MonteCarlo procedure adopted to explore the above
parameter space. We will see that the relevant quantities to be scanned turn out to be 
only the 12 MFV proportionality factors, so that the predictivity and testability of the 
model end up to be dramatically improved.

\section{MFV MSSM predictions for meson mixings} \label{sec:MC}

We are now ready to study the MFV MSSM $\D B = 2$ amplitude, and its predictions for 
$\D M_{s,d}$, by varying the SUSY scales as well as the MFV proportionality
parameters. To this end, a MonteCarlo approach which generates every (or a subset) of the
parameters according to flat distributions within given ranges provides the most
systematic and unprejudiced tool. Here below we describe in more details our adopted
strategy.

\subsection{Strategy} \label{sec:strategy}

Our reference numerical study was carried out by fixing the mass parameters to
`scenarios', and then, for each scenario, scanning with flat distributions the 
12 parameters $(a_i, b_i)$ ruling the MFV expansions (\ref{softMFV}). In addition, 
we set $\tan \be = 3$.\footnote{The impact of variations of $\tan \be$ in the range 
$[3,10]$ was addressed in a specific set of runs to be described below.}

The choice of the mass parameters was designed to cover, in an exhaustive way, 
all the mass scenarios reasonably allowed for low-energy SUSY. To this end, we have 
started from considering the information on the ranges permitted to SUSY masses, 
that is provided by experiments \cite{PDBook}. On this point we make the following remarks
\begin{itemize}

\item Concerning squark masses and $M_{\tilde g}$, the most updated bounds
(\cite{CDF-Ms-Mg,D0-Ms-Mg} and updates thereof) are given in the plane 
$M_{\tilde q} - M_{\tilde g}$, where $M_{\tilde q}$ denotes a generic squark mass 
(see \cite{CDF-Ms-Mg,D0-Ms-Mg}). The profile of the bound is such that when the gluino mass 
can be small, then the generic squark mass is constrained to be large, and viceversa. 
We have chosen four points in the plane $M_{\tilde q} - M_{\tilde g}$, namely
\footnote{The bounds in the $(M_{\tilde q},M_{\tilde g})$ plane provided
by Refs. \cite{CDF-Ms-Mg,D0-Ms-Mg} are in fact somehow tighter than the values chosen in eq.
(\ref{MsqMg}). We note however that these experimental bounds are obtained assuming a
specific mSUGRA scenario. Moreover, in the present study we take the approach of
preferring smaller masses, in order to address the possibility of large signals in meson
mixings. As it will emerge from the discussion, even in this approach NP signals in the 
MFV MSSM are however typically found to be within present errors associated with mixings
themselves.}
\beqn
(M_{\tilde q},M_{\tilde g}) = \{(100,700),(200,500),(300,300),(1000,195)\}~{\rm GeV}~,
\label{MsqMg}
\eeqn
and used them to fix respectively $\ov m$ and $M_{\tilde g}$. We note that $\ov m$ 
is strictly a representative quantity for the squark mass only when the $a$-parameter 
multiplying it (see eq. (\ref{softMFV})) is of O(1). However, the detailed choice 
of $\ov m$ turns out to play a marginal role in our main findings, and the above argument
serves only to give a reasonable criterion on choosing the pair of parameters 
$\ov m$ and $M_{\tilde g}$. We further note that the case in which both $\ov m$ and
$M_{\tilde g}$ are small is `covered' when the $a$-parameter is small.

\item The generic trilinear coupling $A$ was fixed to the value $A = 2 \, \ov m$.
\footnote{For $\ov m = 1000$ GeV we chose also $A = 1000$ GeV.}
This choice, when $\ov m$ is a representative quantity for squark masses, helps 
having a not too low mass for the lightest Higgs \cite{isidori-paradisi}. Again, 
the full spectrum of deviations from this relation is actually covered when scanning 
the $a$-parameters multiplying $\ov m$ and $A$.

\item Constraints on $\mu$ are generically model-dependent. We then considered small,
intermediate and high values for its magnitude by setting the following possibilities
\beqn
\mu = \{\pm 200,\pm 500,\pm 1000\}~{\rm GeV}~.
\label{mu-values}
\eeqn

\item $M_1$ and $M_2$, as well as $\mu$, enter chargino and neutralino mass matrices. 
The choice of $M_1$ and $M_2$ in connection with that of $\mu$ determines the amount 
of gaugino- and higgsino-like components in their field content. In order to have 
representative cases with respect to the experimental information \cite{PDBook}, we made 
the following choices
\beqn
\begin{tabular}{lcl}
$|\mu| = 200$ & $\Rightarrow$ & $(M_1,M_2)=\{(500,500),(1000,1000)\}$ GeV~,\\
$|\mu| = 500$ & $\Rightarrow$ & $(M_1,M_2)=\{(100,200),(500,500)\}$ GeV~,\\
$|\mu| = 1000$ & $\Rightarrow$ & $(M_1,M_2)=\{(100,200),(100,500)\}$ GeV~,
\end{tabular}
\label{M1M2-values}
\eeqn
\ie two possible choices for every of the six $\mu$ values listed in eq. (\ref{mu-values}). 
Choices (\ref{M1M2-values}) translate into values for the masses of the lightest 
chargino and neutralino, which in turn tune the importance of the respective box
contributions. We note that, in our case, neutralinos have almost no impact on the sum of the
contributions, even when they are very light. This applies also to the mixed 
gluino-neutralino boxes. In this respect, we observe that, from eqs.
(\ref{MsqMg})-(\ref{M1M2-values}), there are unphysical cases among our considered scenarios
in which the lightest SUSY particle (LSP) is not a neutralino. However, given the mentioned 
marginal impact of neutralino masses on our main findings, one can always lower the value of 
$M_1$ in order to have a neutralino as the LSP. The values chosen in eq. (\ref{M1M2-values}) 
are intended to ascertain that the impact of the choice of the neutralino masses be in fact minimal.\\
Concerning charginos, their contribution is regulated by the
lightest between $M_2$ and $|\mu|$ (see eq. (\ref{MCha})), the detailed choice of the
other parameters playing basically no role on the main findings we will discuss.

\item The remaining two parameters $m_{H_{u,d}}$ enter exclusively Higgs boxes. 
As we also remarked at the end of Section \ref{sec:MSSM-DF2}, the calculation of 
the latter can be `factored out', since they depend only on $m_{H_{u,d}}$ 
and on $\mu$, through relation (\ref{MH1}), and on no other SUSY scale. 
With reference still to the discussion at the end of Section \ref{sec:MSSM-DF2}, it is
also clear that the single relevant SUSY scale introduced by the Higgs sector is
the physical charged Higgs mass $M_{H_1^{\pm}}$, eq. (\ref{MH1}), and not $m_{H_{u,d}}$ 
separately.\\
The dependence of Higgs contributions on variations of $M_{H_1^{\pm}}$ and on $\tan \be$ 
will be studied in a separate Section below. Here we mention that such contributions are 
positive for every allowed value of $M_{H_1^{\pm}}$ if $\tan \be \le 7$, and even for 
$\tan \be = 10$, they reach (small) negative values only with very light 
$M_{H_1^{\pm}}$. As a consequence, their impact for low $\tan \be$ is just an overall 
(positive) shift of the sum of the other contributions.

\end{itemize}

This completes the discussion on the choice of the SUSY scales in our main analysis. 
The mass scenarios explored amount to 48. 
Taking into account the various remarks made above on every specific subset 
of the parameters, we believe that such analysis covers extensively all the 
interesting combinations in the SUSY parameter space.

For each of the above scenarios, we then scanned the MFV parameters $a_i,b_i$ 
assuming (uncorrelated) flat distributions according to (see also footnote \ref{b3b4})
\beqn
0.25 \le a_{1,2,3} \le 1~, ~~~ -1 \le \{a_{4,5}, b_{1,...,8}\} \le 1~.
\label{aibi-values}
\eeqn
The lower bound in $a_{1,2,3}$ was not chosen to be zero, in order not to have 
to discard most of the resulting squark matrices because of a negative lowest 
eigenvalue.

We finally note that, in our analysis, we do not include other FCNC constraints which
could in principle play a role for low $\tan \beta$, in particular $b \to s \gamma$. In
this respect we observe that, as already mentioned in the introduction, NP corrections to
meson mixings within the MFV MSSM are typically within present errors and the
inclusion of additional constraints can only further suppress NP signals. In addition, 
as again mentioned in the introduction, our analysis will lead to the identification of mass 
regimes, ruled by the interplay between chargino and gluino contributions, with the Higgs
contributions discussed separately. Since in $b \to s \gamma$ the main role is played by
chargino and Higgs contributions (see e.g. \cite{CDGG}), it is clear that the $b \to s \gamma$ 
constraint would not exclude any of the above regimes.

\subsection{Results} \label{sec:results}

We now discuss our results. The latter were all obtained using the MonteCarlo strategy
outlined in the previous discussion. We have however also verified the specific findings
with alternative runs, designed to uncover possible loopholes. We will refer to them in
due course.

Our phenomenological analysis starts from the calculated meson-antimeson oscillation 
amplitude
\beqn
\mc M^{(M)}_{12} \equiv \< M | \heff | \ov M \>~,
\label{M12}
\eeqn
with $M = K$, $B_{d,s}$. Wilson coefficients, evaluated at the matching scale, are 
subsequently run to the $m_b$ pole mass or to $2$ GeV, the scales at which the effective 
matrix elements are evaluated on the lattice \cite{Bparams-lattice-DB,Bparams-lattice-DS} in 
the $B_{d,s}$ and $K$ case, respectively (see also
\cite{DB=2-lat1,DB=2-lat2,DB=2-lat3,DB=2-lat4,Dalgic}).
In the running, we used NLO formulae from \cite{NLOADM,NLOADMcheck}, with the matching
scale chosen to be at $350$ GeV, as a compromise between all the SUSY scales entering the 
calculation.\footnote{Variations around this value have basically no effect on the results.}

Then, from twice the amplitude (\ref{M12}), one can calculate the experimentally measured 
mass differences by taking the absolute value or respectively the real part in the
$B_{d,s}$ or $K$ cases \cite{BurasLH}. In the present study, we restrict to $\D M_{s,d}$.

\begin{figure}[th!]
\begin{center}
\includegraphics[scale=0.50]{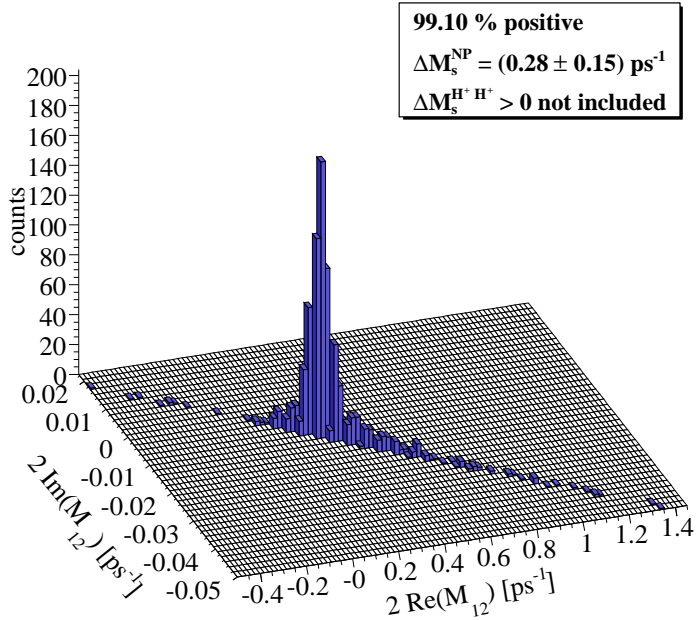}
\end{center}
\caption{\small\sl Lego plot showing the alignment of the phase of SUSY contributions to
the SM phase in the MFV MSSM $\mc M_{12}$ for the case of $B_s$.
In this example SUSY scales are (GeV): $\mu=1000$, $\ov m=200$, $M_{\tilde g}=500$,
$M_1=100$, $M_2=500$. In the legend, $\D M_s$ is calculated from the absolute value
formula, keeping the sign of the real part. The percentage gives the integrated number of
hits for which $\D M_s^{\rm NP}>0$.}
\label{fig:h2sum3D}
\end{figure}

As a first check, we have verified that the phase of meson-antimeson oscillations in the 
MFV MSSM be aligned with the SM one. This feature is shown in Fig. \ref{fig:h2sum3D} for
the $\D M_s$ case. In the Figure, we have chosen a specific mass scenario for the SUSY
scales and scanned the MFV parameters $a_i$, $b_i$, obtaining a distribution of
values for $\mc M_{12}$. As expected, values are aligned along a line with slope 
$\tan(\arg \mc M_{12})$.

In MFV, the phase of meson-antimeson oscillations is by definition not a good observable
to search for NP effects. However, the same does not apply to the mass differences. As a
matter of fact, by studying the latter, we found a number of interesting and sometimes
surprising features, which we now discuss.

\subsubsection*{i. NP contributions are positive}

\begin{figure}[t]
\begin{center}
\includegraphics[scale=0.35]{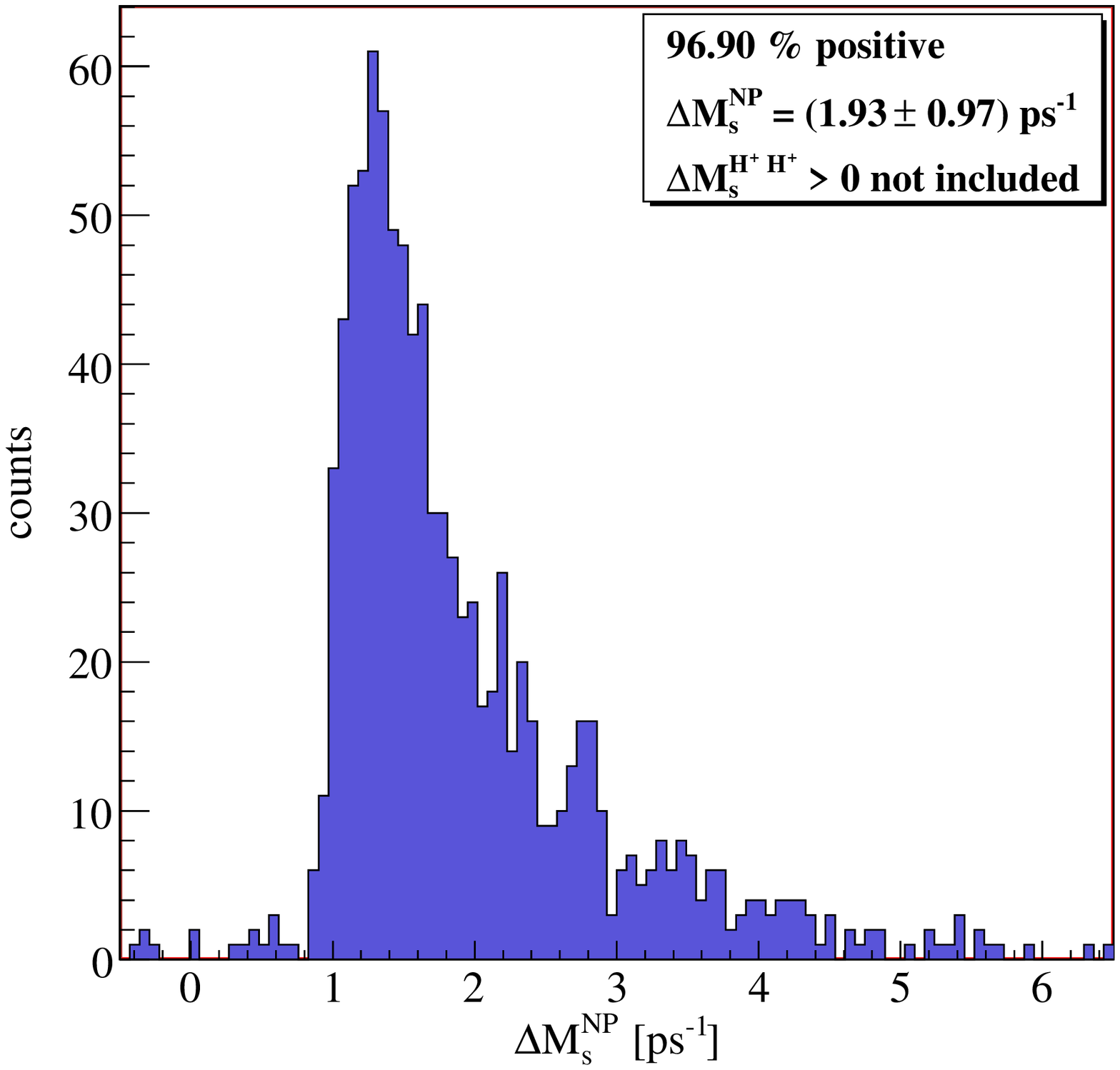}
\includegraphics[scale=0.35]{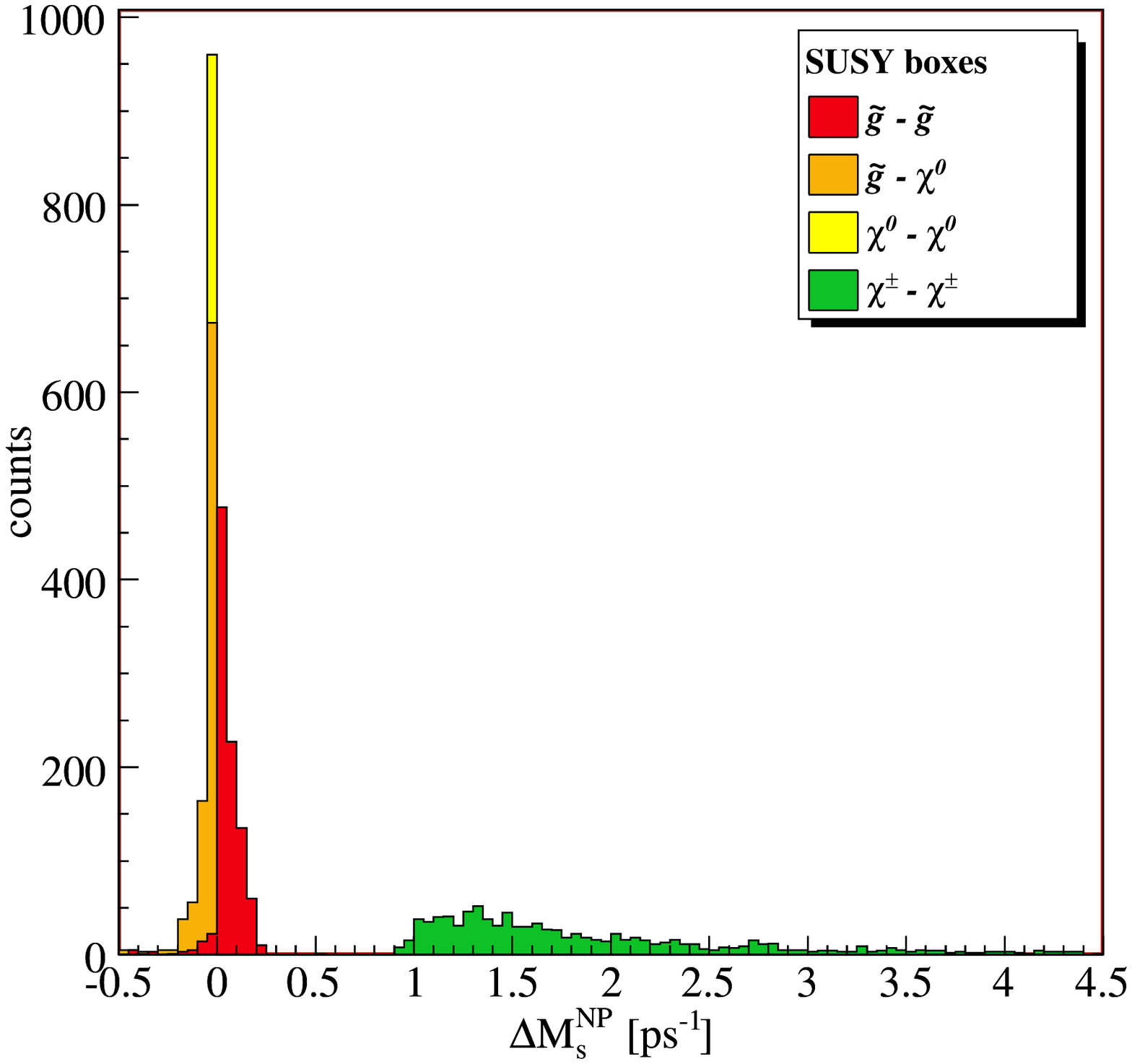}
\end{center}
\caption{\small\sl Distribution of values for $\D M_s^{\rm NP}$ in the MFV MSSM: sum of
the contributions (left panel) and separate SUSY contributions (right panel).
The distribution results from scanning the MFV parameters $a_i$, $b_i$, after choosing 
SUSY scales as (GeV): $\mu=200$, $\ov m=300$, $M_{\tilde g}=300$, $M_1=500$, $M_2=500$. 
In the plot and in the legend, $\D M_s$ is calculated from the absolute value formula, keeping 
the sign of the real part. The percentage gives the integrated number of hits for which 
$\D M_s^{\rm NP}>0$. [See also text, mass regime (A).]}
\label{fig:sum-mu200}
\end{figure}

\begin{figure}[t]
\begin{center}
\includegraphics[scale=0.35]{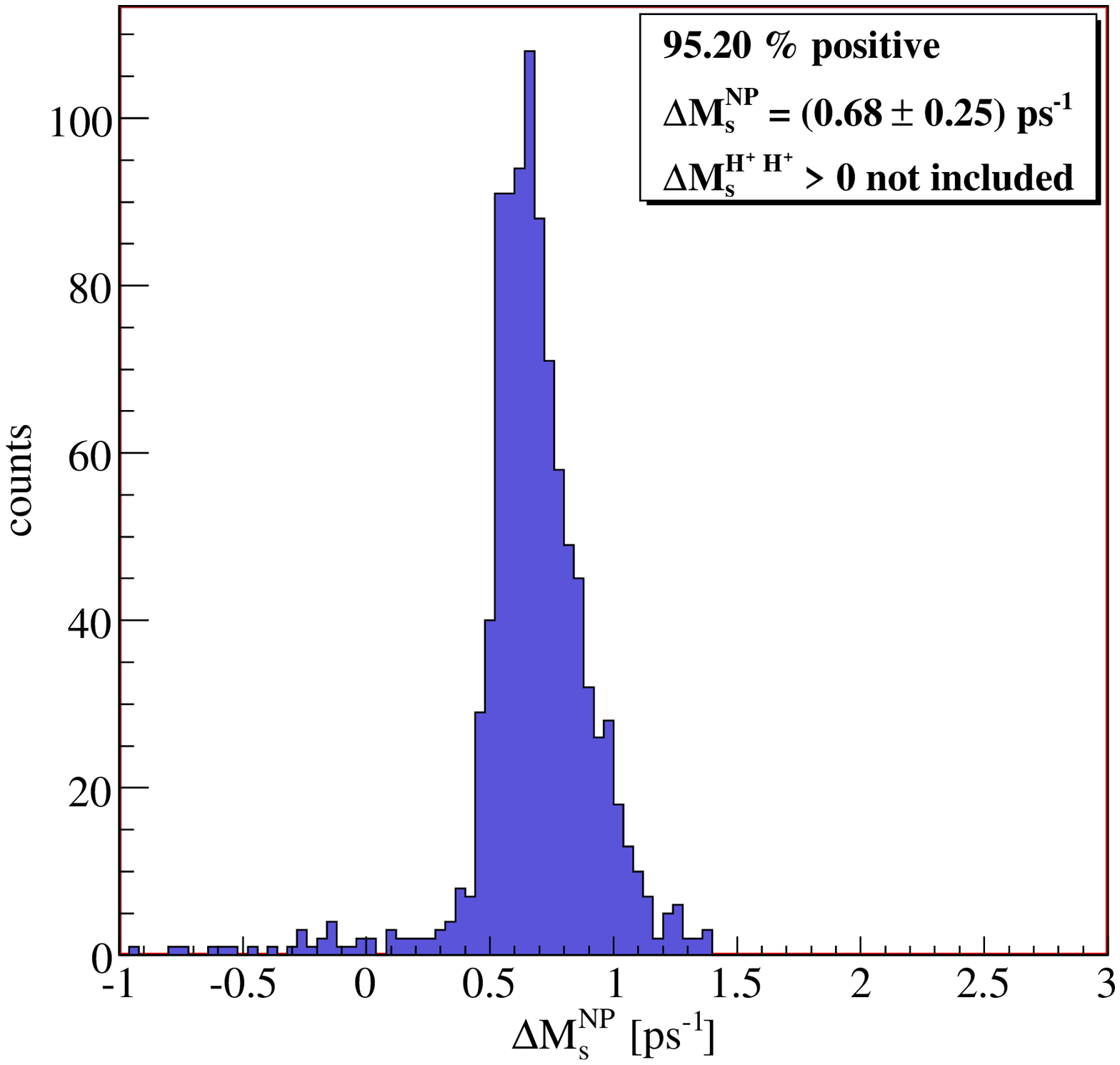}
\includegraphics[scale=0.35]{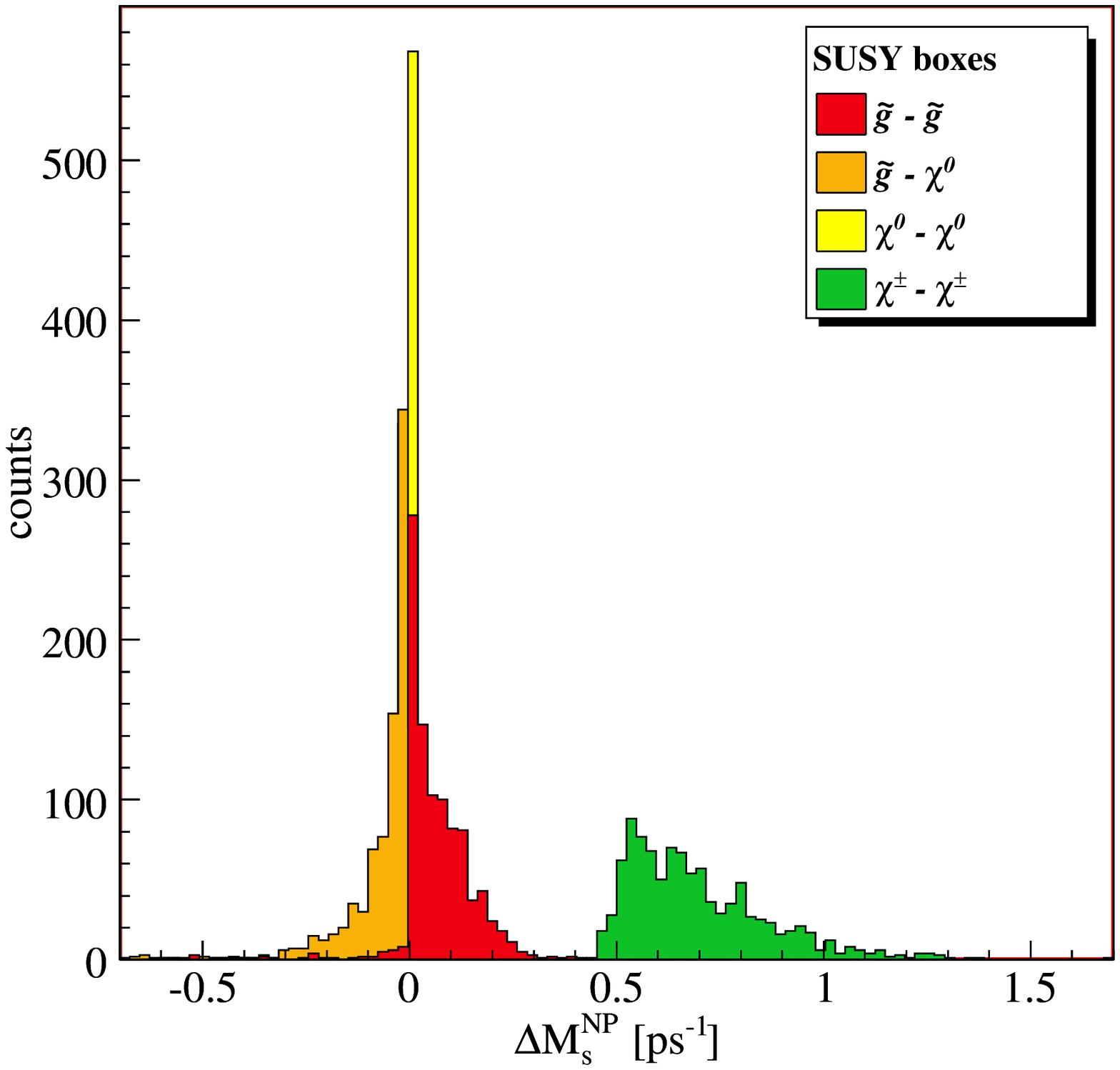}
\end{center}
\caption{\small\sl Same as Fig. \ref{fig:sum-mu200} but for SUSY scales chosen as (GeV): 
$\mu=500$, $\ov m=300$, $M_{\tilde g}=300$, $M_1=100$, $M_2=200$. [See also text, mass
regime (B).]}
\label{fig:sum-mu500}
\end{figure}

\begin{figure}[t]
\begin{center}
\includegraphics[scale=0.35]{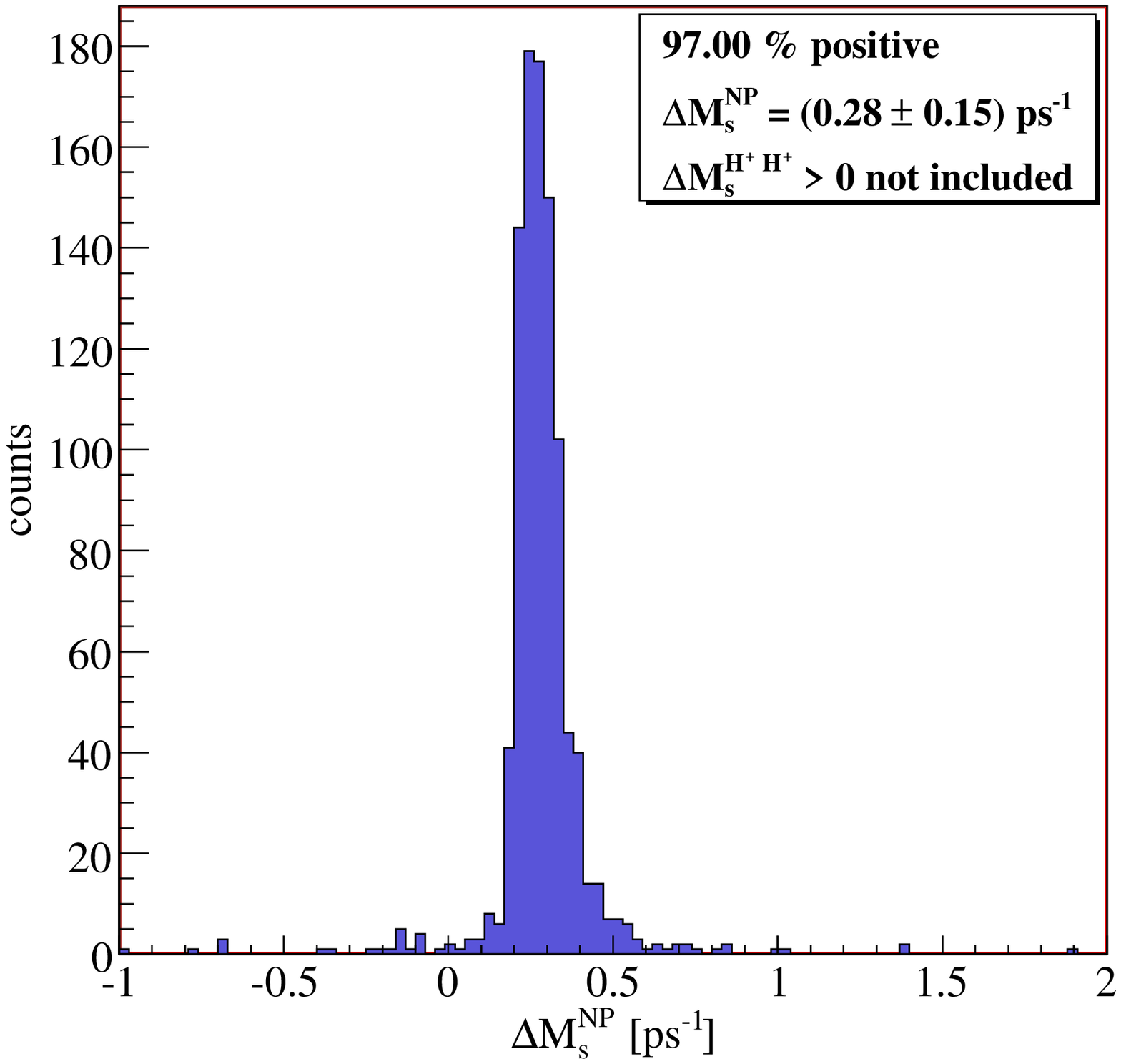}
\includegraphics[scale=0.35]{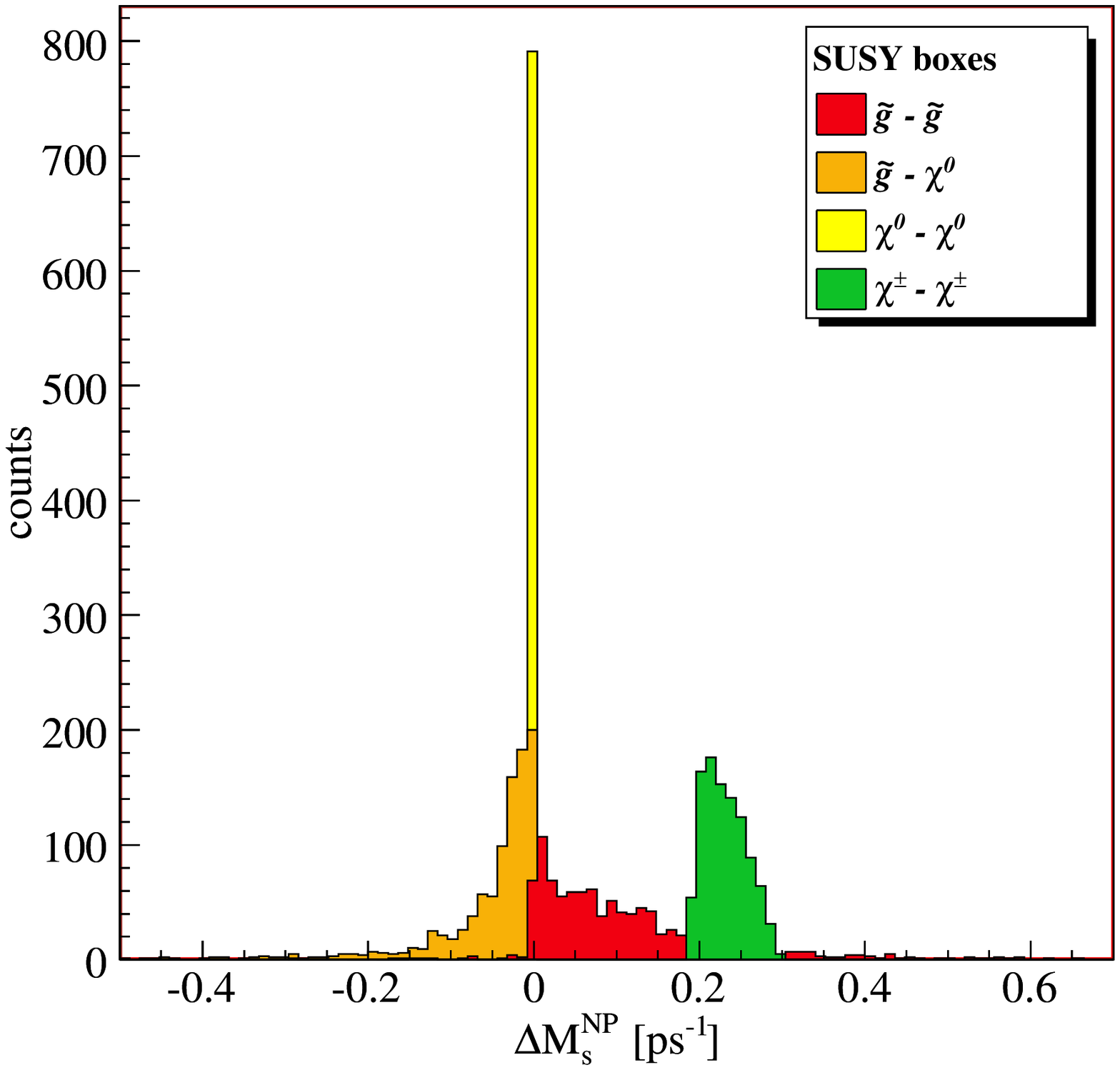}
\end{center}
\caption{\small\sl Same as Fig. \ref{fig:sum-mu200} but for SUSY scales chosen as (GeV): 
$\mu=1000$, $\ov m=300$, $M_{\tilde g}=300$, $M_1=100$, $M_2=500$. [See also text, mass
regime (C).]}
\label{fig:sum-mu1000}
\end{figure}

\begin{figure}[t]
\begin{center}
\includegraphics[scale=0.35]{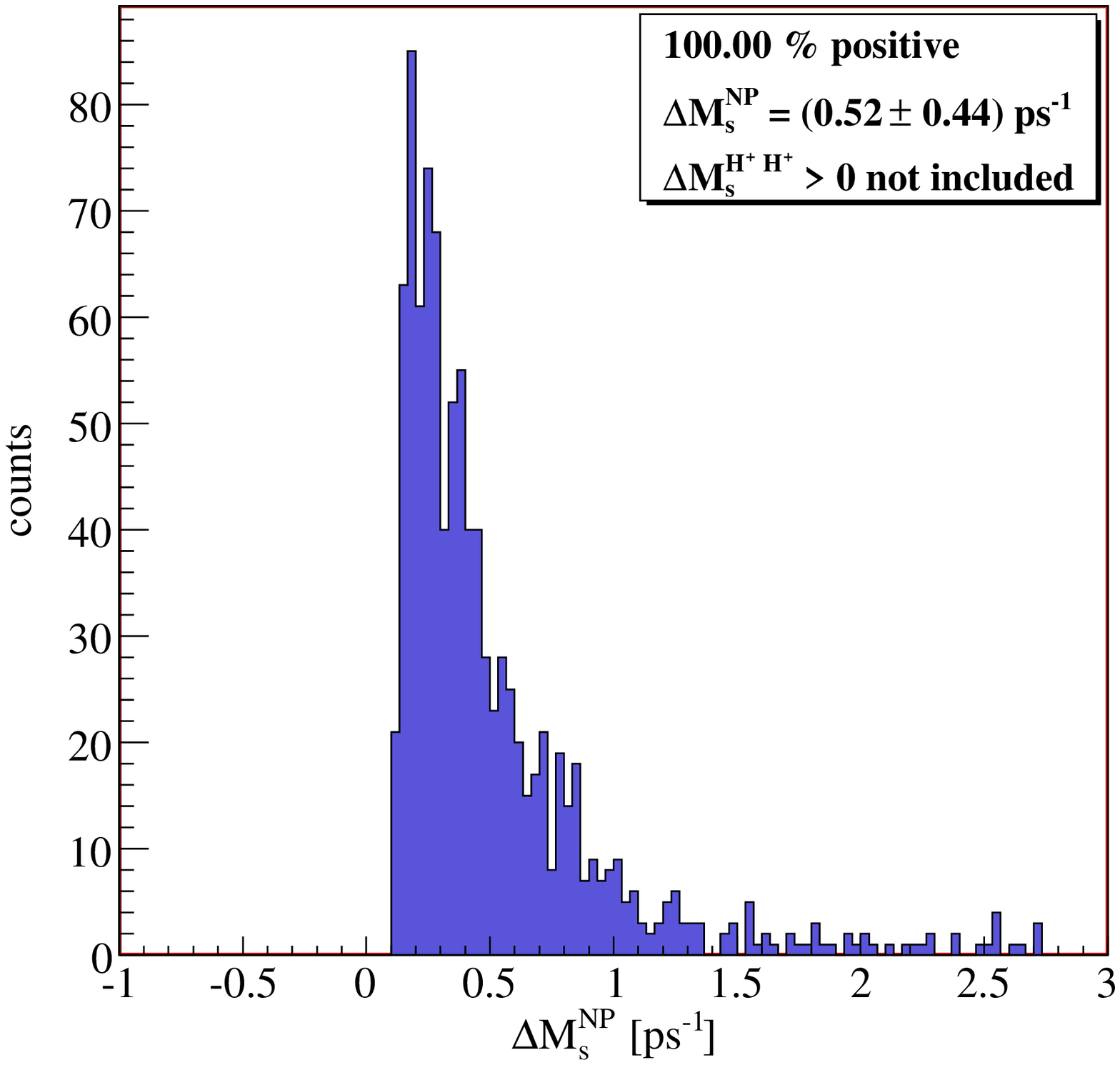}
\includegraphics[scale=0.35]{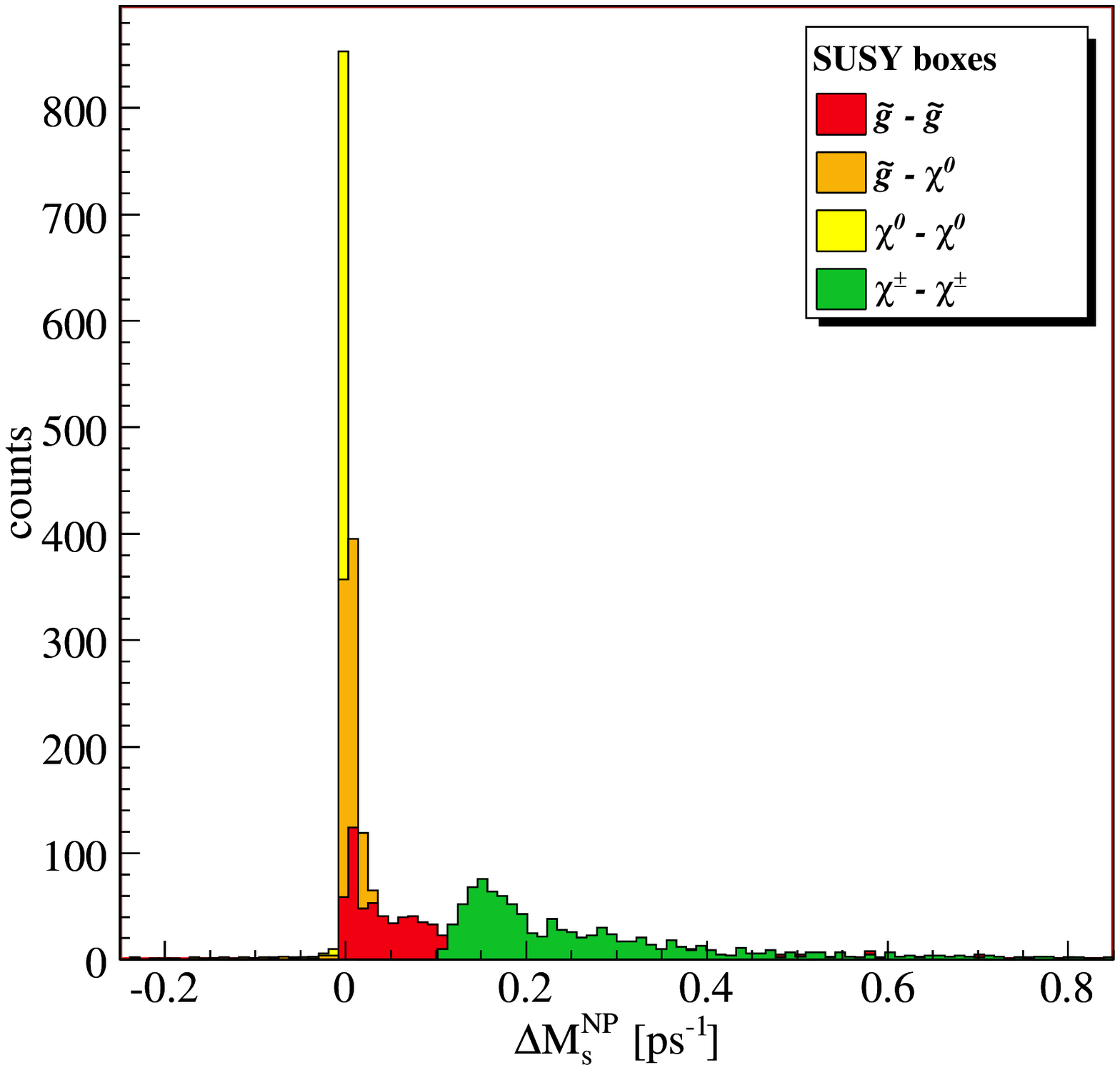}
\end{center}
\caption{\small\sl Same as Fig. \ref{fig:sum-mu200} but for SUSY scales chosen as (GeV): 
$\mu=500$, $\ov m=1000$, $M_{\tilde g}=195$, $M_1=100$, $M_2=200$. [See also text, mass
regime (D).]}
\label{fig:sum-bigMsq}
\end{figure}

A first surprising fact emerges by studying the sum of the SUSY contributions to the meson
mass differences. As already mentioned above, for every of the mass scenarios considered,
we have randomly generated the MFV parameters $a_i$, $b_i$. The obtained distribution of values 
in $\mc M_{12}$ translates into a corresponding distribution for the meson mass
differences. As an example, one can look again at Fig. \ref{fig:h2sum3D}, displaying $\mc
M_{12}$ for the $B_s$-meson. In this case, ${\rm Im}(\mc M_{12}) \cong 0$ and an excellent 
estimate of $\D M_s^{\rm NP}$ is provided by the projection of the distribution along the 
${\rm Re}(\mc M_{12})$. In the left panels of Figs. \ref{fig:sum-mu200} to
\ref{fig:sum-bigMsq} we show how the distribution in $\D M_s^{\rm NP}$ looks like in four
representative scenarios.\footnote{In the plots, the number of `events' obtained after
scanning $a_i$, $b_i$ is set to 1000. We have verified that the distributions are
left qualitatively unchanged when considering subsets of these 1000 points and when changing 
the binning, so that 1000 is a statistically significant number.}

As one can immediately realize, almost the totality of points features $\D M_s^{\rm NP}>0$, 
\ie SUSY contributions to $\D M_s$ in the MFV MSSM at low $\tan \be$ are positive. We
explicitly mention that in Fig. \ref{fig:h2sum3D}, as well as in Figures
\ref{fig:sum-mu200}-\ref{fig:sum-bigMsq} below, charged Higgs boxes are not included. Their
contribution amounts to a further {\em positive} shift of the distribution representing
the sum of the other contributions, since Higgs boxes do not depend on the MFV parameters 
$a_i, b_i$.

In the following discussion, we will get further insight on the positiveness of the SUSY
corrections to meson mixings, by analyzing the separate SUSY contributions which sum up 
to give $\D M_s^{\rm NP}$. Their interplay and a number of additional checks turn out to 
provide as many arguments in support of the above statement.

\subsubsection*{ii. Mass regimes}

The positiveness of the sum of SUSY contributions holds true, irrespective of the mass 
scenario chosen. As a matter of fact, the 48 scenarios we have considered turn out to be
classifiable into 4 main `mass regimes', each characterized by a definite interplay among the
various SUSY contributions. The deciding factors are the choice of the generic squark mass 
$\ov m$ and of the magnitude of $\mu$, the rest of the mass parameters, as well as the
sign of $\mu$, playing only a minor role once these are fixed. Specifically, one can choose a 
`not large' value for the generic squark scale $\ov m$ (\ie $\ov m < 1000$ GeV in our 
eq. (\ref{MsqMg})) {\em and} {\bf (A)} $|\mu|$ small or {\bf (B)} $|\mu|$ intermediate or 
{\bf (C)} $|\mu|$ large. Alternatively, there is one more mass regime {\bf (D)} when one 
chooses a large value for $\ov m$, irrespective of the choice of all the other SUSY scales, 
including $\mu$. The four scenarios displayed in Figs. \ref{fig:sum-mu200} to 
\ref{fig:sum-bigMsq} are representative of such mass regimes, in the order {\bf (A)} to 
{\bf (D)}.

\subsubsection*{iii. Interplay between chargino and gluino boxes}

Let us now have a closer look into the various mass regimes, by discussing, within each 
of them, the main features of the separate contributions. The latter are displayed in the
right panels of Figs. \ref{fig:sum-mu200} to \ref{fig:sum-bigMsq}. As one can see from the
figures -- and as it will emerge from the subsequent discussion -- the main actors in
determining the sum of SUSY contributions to meson oscillations are chargino and gluino
boxes, the remaining ones playing a minor role. We remind the reader that Higgs 
contributions are not considered in this discussion and not included in 
Figs. \ref{fig:sum-mu200} to \ref{fig:sum-bigMsq}. For low $\tan \be$, Higgs contributions 
trivially amount to a positive shift of the total result. We will address this point in a 
specific section.

{\bf Mass regime (A)}. Here, the smallness of $\mu$ implies a low value for the lightest
chargino mass eigenstate (see eq. (\ref{MCha})). As a consequence, contributions from
charginos tend to be large, of the order of $2$ ps$^{-1}$. In addition they are positive, 
being dominated by the SM operator $\mc Q_1$ \cite{gabrielli-giudice}. Gluino contributions 
are negligible in this scenario, and the reason, still related with the smallness of $\mu$,
will be clear from the discussion in mass regime {\bf (C)} below.\\
A decrease in $\ov m$ only reinforces the chargino dominance. One finds in this case that 
gluino boxes remain negligible, while, for chargino ones, the up-squark mass scale gets 
lower and their contribution is correspondingly increased. As a matter of fact, 
Fig. \ref{fig:sum-mu200} shows somehow the `worst' case among the low $|\mu|$ ones within 
our studied scenarios. In the other cases, the chargino dominance is even more evident 
and the number of points with $\D M_s^{\rm NP}>0$ even closer to $100$ \%.

{\bf Mass regime (B)}, with $|\mu|$ moderate, is a case of transition, intended to show
the rate of decrease in importance of chargino contributions with increasing $|\mu|$. 
Chargino contributions are still dominant as compared to gluino ones. Hence, in this respect, 
the situation is not qualitatively different from regime {\bf (A)}. 
On the whole, an increase in $\mu$ from $200$ to
$500$ GeV corresponds to a decrease in the total signal from $\approx 2$ to $\approx 1$
ps$^{-1}$, and similarly to regime {\bf (A)}, the total signal drops abruptly beneath the
peak value.

{\bf Mass regime (C)} occurs then for large $|\mu|$. In this case, the flavour diagonal LR 
entries in the down-squark mass matrix are large (see eq. (\ref{Md})) and in connection with
the flavour mixing induced by LL entries, enhance contributions from scalar operators in 
gluino boxes. Even if the up-squark mass matrix has in principle the same structure, the 
same mechanism turns out to be not efficient in enhancing chargino
scalar contributions.\footnote{One can provide an intuitive argument for this fact as follows:
gluino contributions to the various operators have the structure 
$Z_D Z_D ({\rm loop~function}) Z_D^\dagger Z_D^\dagger$, with $Z_D$ defined in eq.
(\ref{Zud}). Such structure holds for every operator. On the other hand, chargino
contributions to scalar operators have the structure $V_L V_R ({\rm loop~function})
V_L^\dagger V_R^\dagger$, with $V_{L(R)}$ the left (right) chargino-up squark-down quark
vertex coupling (see Appendix), while a similar structure -- but with four $V_L^{(\dagger)}$ 
vertices -- holds for the contributions to $\mc Q_1$. Now, since $V_L \sim Y_u$ and 
$V_R \sim Y_d$ (see Appendix \ref{app:XN-couplings}), in the case of charginos, 
contributions other than $\mc Q_1$ are always made small by a suppression factor of $(Y_d/Y_u)^2$.}
The main parametric dependence ruling the above mechanism for gluino contributions is then
the product $\mu \times \tan \beta$ (see eq. (\ref{Md})). One should also note that
the increase in the value of $|\mu|$ with respect to the previous regimes largely
suppresses chargino contributions, helping in turn chargino and gluino boxes to become
of comparable size.\\
The right panel of Fig. \ref{fig:sum-mu1000} displays a typical case for mass regime (C),
with gluino contributions amounting to roughly $30 \div 50$ \% positive corrections to the 
chargino signal. We note here the occurrence of another interesting mechanism: in cases
where the squark scale $\ov m$ and/or the mass $M_1$ are small (as in the example of 
Fig. \ref{fig:sum-mu1000}), gluino-neutralino boxes give a {\rm negative} and relatively
important contribution. However, the latter is typically outpaced by the positive
contribution from pure gluino boxes (plus of course that from charginos), with a total
signal around $0.5$ ps$^{-1}$. A similar mechanism can already be recognized in regime 
(B) (see Fig. \ref{fig:sum-mu500}), but in that case is less evident.\\
We mention that, within the set of scenarios corresponding to regime (C), we found `extreme'
cases where scalar contributions from gluino boxes completely overwhelm any other contribution. 
These occur when choosing a very light squark scale, $\ov m = 100$ GeV. In these cases, chargino 
boxes amount to a small positive signal, while gluino-neutralino boxes give a contribution which 
is negative and relatively large. The latter is again significantly counterbalanced by the positive, large 
signal from gluinos and the sum of contributions results in a positive, quite spread signal for 
$\D M_s^{\rm NP}$. However, such extreme cases correspond to very light squark masses, the lightest 
down- and up-squarks being around $30$ and $60$ GeV, respectively, which is very 
unrealistic.\footnote{As a further remark, we note that choosing $|\mu|$ large, with $\ov m$ small, 
causes the squark mass matrices' determinants to be negative for most of the parameter space in the 
MFV parameters $a_i,b_i$, implying in turn an odd number of negative mass squared
eigenvalues. The corresponding point in the parameter space is then discarded as unphysical. As 
a matter of fact, for $|\mu| = 1000$ GeV and $\ov m = 100$ GeV, the ratio of discarded to
valid points is $\approx 125$, but this number drops to $\simle 2$ already for $\ov m = 200$ GeV.}

{\bf  Mass regime (D)} is characterized by a large value for the squark
scale $\ov m$, in our case $1000$ GeV, and basically unaffected by the choice of any other
parameter. In this case, the largeness of the squark scale sets to zero gluino-neutralino
contributions, whose negative skewness had some effect in the previous cases. 
On the other hand, both chargino and gluino distributions (positive) are characterized 
by a long tail, as shown on the right panel of Fig. \ref{fig:sum-bigMsq}. The respective 
contributions are comparable in size, with the magnitude of those from gluinos growing 
with $|\mu|$. The average SUSY signal is generically small, $\simle 1$ ps$^{-1}$.

\bigskip

As a last overall remark, we explicitly note that, in all regimes considered, the magnitude
of contributions coming from the flavour off-diagonal elements in the down-squark matrix,
entering gluino and neutralino boxes, typically does not exceed $0.5$ ps$^{-1}$.
Therefore the latter set of contributions is strictly important only when also chargino 
contributions are small, \ie in regimes (C) and in particular (D). We may add that such
regimes are phenomenologically relevant after the simple observation that the experimental 
measurement of $B_s$ oscillations \cite{CDF-DMs} and its agreement with the SM central value 
undoubtedly favour small NP corrections with respect to large ones. The final word will be 
provided by a substantial decrease of the lattice error, to the level of a few percent.

\subsubsection*{iv. Role of charged Higgs boxes}

In the above study, we have completely omitted the inclusion of charged Higgs boxes.
Recalling the discussion at the end of Section \ref{sec:MSSM-DF2}, we can now investigate 
their contribution separately, as a function of the single new SUSY scale they introduce, 
namely the physical charged Higgs mass $M_{H^{\pm}_1}$.

\begin{figure}[t]
\begin{center}
\includegraphics[scale=0.50]{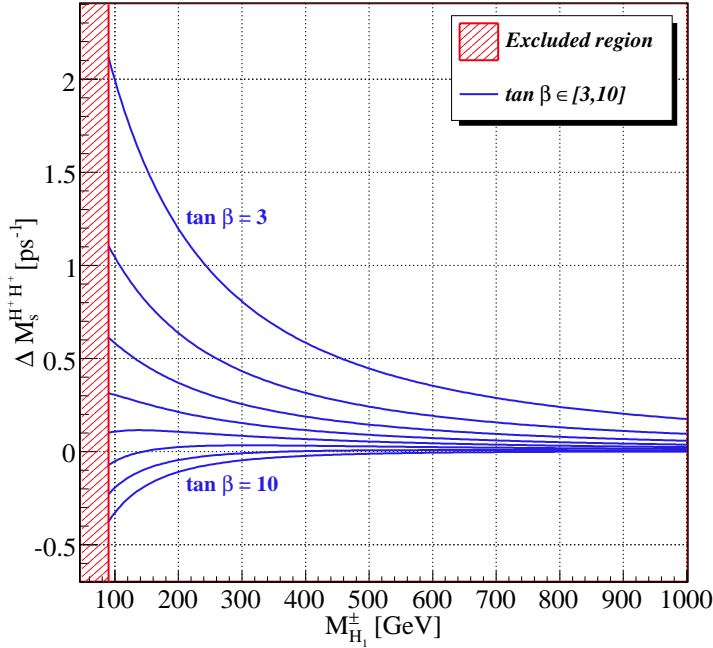}
\end{center}
\caption{\small\sl Charged Higgs boxes contribution to $\D M_s$ as a function of the
physical charged Higgs mass $M_{H^{\pm}_1}$. The different curves correspond to increasing
values of $\tan \be$ from $3$ (uppermost one) to $10$ (lowermost one). The hatched area on
the left of $M_{H^{\pm}_1} = 90$ GeV represents the experimentally excluded region
\cite{PDBook} from direct search only.}
\label{fig:Higgs}
\end{figure}

In Fig. \ref{fig:Higgs} we report the charged Higgs contribution to $\D M_s$ as a function 
of $M_{H^{\pm}_1}$, for values of $\tan \be$ between $3$ and $10$. The hatched area 
on the left of $M_{H^{\pm}_1} \cong 90$ GeV represents the region excluded after direct 
experimental searches \cite{PDBook}.

As one can see, for $\tan \be \le 7$ charged Higgs boxes give a positive correction to 
$\D M_s$, irrespective of the value assumed by the physical mass $M_{H^{\pm}_1}$. 
For this reason, we have chosen not to include such correction in the previous study
on our mass scenarios: for each of them, Higgs contributions amount to a constant,
positive shift of the final result. Then, in particular, the plots of Figs. 
\ref{fig:sum-mu200}-\ref{fig:sum-bigMsq} provide only a `lower bound' on $\D M_s^{\rm NP}$, 
to be augmented by the contribution from Higgs boxes.

It is interesting to give instead an indicative upper bound on the MFV MSSM contributions
to $\D M_s^{\rm NP}$ for low $\tan \be$. To this aim, we can consider mass regime (A) --
chargino dominated -- which tends to give the largest positive corrections (around $2$
ps$^{-1}$, see Fig. \ref{fig:sum-mu200}), together with Higgs contributions taken at a
small value for $M_{H^{\pm}_1} \approx 100$ GeV. By looking at Fig. \ref{fig:Higgs}, one
can see that Higgs corrections amount to roughly another $2$ ps$^{-1}$. So, one can
estimate MFV MSSM corrections to $\D M_{s}$ not to exceed $4$ ps$^{-1}$. However, they are
typically considerably smaller than this upper bound, as one can see by inspection of the
various mass regimes (see also Fig. \ref{fig:runtb} below). This fact shows, for the case 
of $B_s - \ov B_s$ mixing, that explicit implementation of the MFV limit in the MSSM leads 
to a naturally small correction to the SM prediction: in fact, expansions (\ref{softMFV}) 
analytically realize the condition of ``naturalness of near flavour conservation''
advocated in \cite{hall-randall-MFV}.

\subsubsection*{\boldmath v. The case of $\D M_d$}

We applied the entire strategy described above also to the case of $\D M_d$. Results are
completely analogous, so we will limit to a few observations.

Keeping for the moment aside charged Higgs contributions, the magnitude of SUSY corrections 
$\D M_d^{\rm NP}$, normalized to the SM prediction $\D M_d^{\rm SM}$, is basically the same 
as the corresponding quantity in the $B_s$ case, in all our studied scenarios. Also the
shapes of the distributions of values for the single contributions look very similar in the $B_d$ 
and $B_s$ cases.

This leads to the following remark. Even in scenarios where gluinos give important contributions
from scalar operators, the latter do not bring about a sensible dependence on the external
quark flavours. In fact, leading scalar contributions go as $\sim m_b^2$, and those in 
$\sim m_b m_s$ or respectively $\sim m_b m_d$ are subleading ones. The effect of the
latter is moreover completely hidden, in our case, by the lack of knowledge of the SUSY
scales and of the MFV parameters.

The inclusion of charged Higgs contributions does instead `distinguish' the case of 
$\D M_d$ with respect to that of $\D M_s$. Now contributions to the LR scalar operator
behave as $\sim m_b m_{s (d)} \, \tan^2 \be$ for $\D M_{s (d)}$ (look at the coefficient
$C_4^{H^+ H^+}$ in the Appendix).\footnote{At the same time, contributions to $C_1^{H^+
H^+}$ are suppressed by $\cot^2 \be$.} For $\tan \be < 10$ , however, effects on the ratio $\D
M_d / \D M_s$ are within $1$ \%. Only for $\tan \be \ge 10$ do they become larger than $3$
\%, and can be visible once the lattice error on $\xi$ is at least halved with respect to
the present value.

\section{\boldmath Additional MonteCarlo's and role of $\tan \be$} \label{sec:discussion}

In this Section we elaborate on our main findings, as described in the previous study. In
particular, we provide further arguments in support of our results, by verifying them
with a set of additional numerical studies. The latter are designed to explore possible 
loopholes in the above treatment, and to address the question how the picture changes when 
increasing $\tan \be$ from strictly low values.

From the above discussion, it is evident that the positiveness of the SUSY correction 
$\D M_s^{\rm NP}$ in the MFV MSSM is due to an interplay among different contributions,
the most important being those from charginos and gluinos. In addition, for 
$\tan \be \le 7$, Higgs boxes further shift the result by a positive amount, depending on 
the chosen $M_{H^{\pm}_1}$.

Such findings followed from a MonteCarlo study in which SUSY masses were fixed and MFV
parameters generated with flat distributions. To check for the robustness of our results,
we also performed additional MonteCarlo's.

\subsubsection*{$\bullet$ Random scan of both SUSY scales and MFV parameters}

\noindent In a first alternative set of runs, we scanned both SUSY scales 
and MFV parameters. In particular SUSY masses were assumed to follow flat distributions in 
the range $M_i \in [100,1000]$ GeV.\footnote{The lower bounds for $M_{\tilde g}$ and for 
$M_{H^{\pm}_1}$ were set to $200$ GeV and $90$ GeV, respectively.} 
In such runs, we also set $\tan \be$ to different values in the range $\tan \be \in [3,15]$. 
This allowed us to study the resulting modifications in the various contributions, in 
particular in those from Higgs boxes, which for a light $M_{H^{\pm}_1}$ tend to be negative 
when $\tan \be > 7$.

\begin{figure}[th!]
\begin{center}
\includegraphics[scale=0.35]{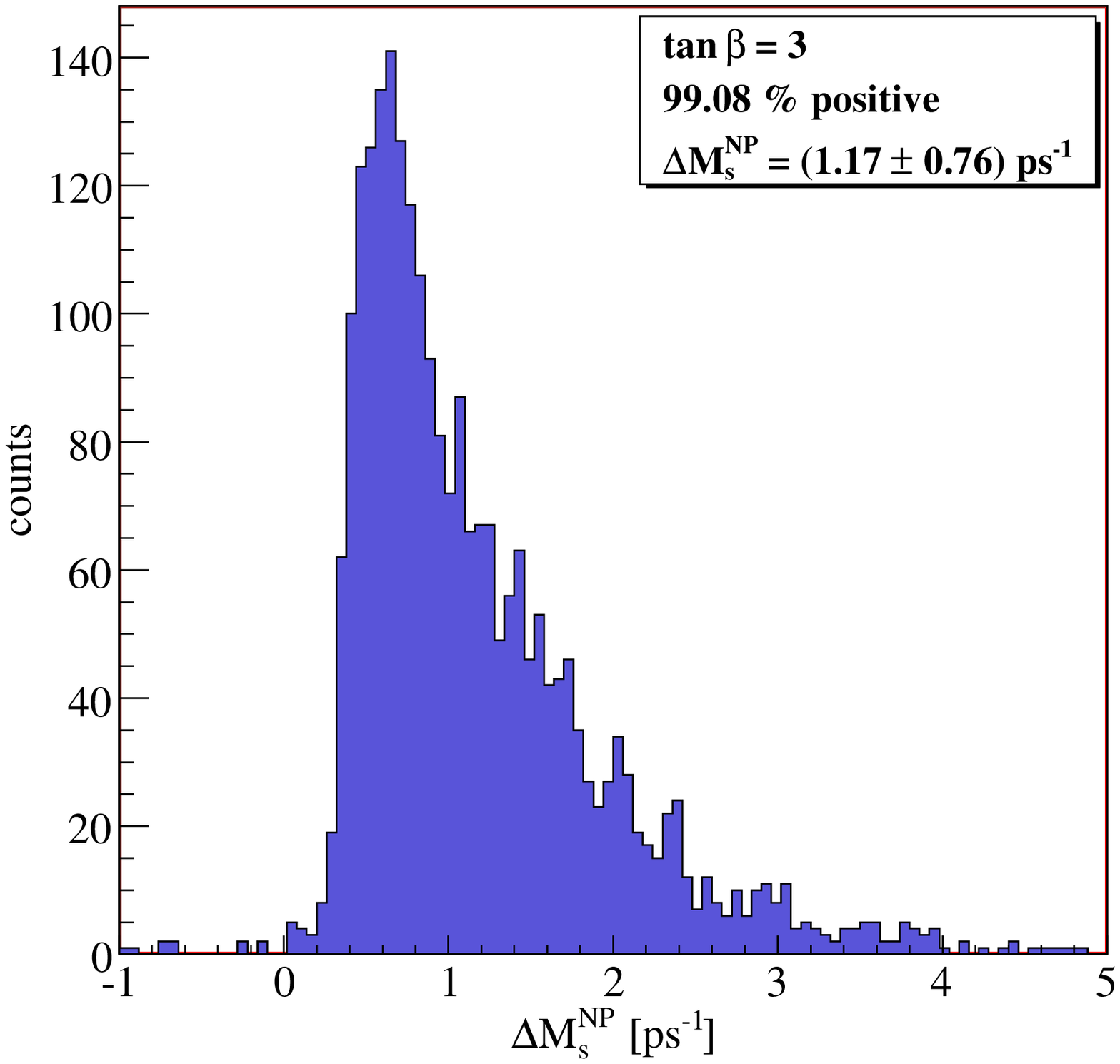}
\includegraphics[scale=0.35]{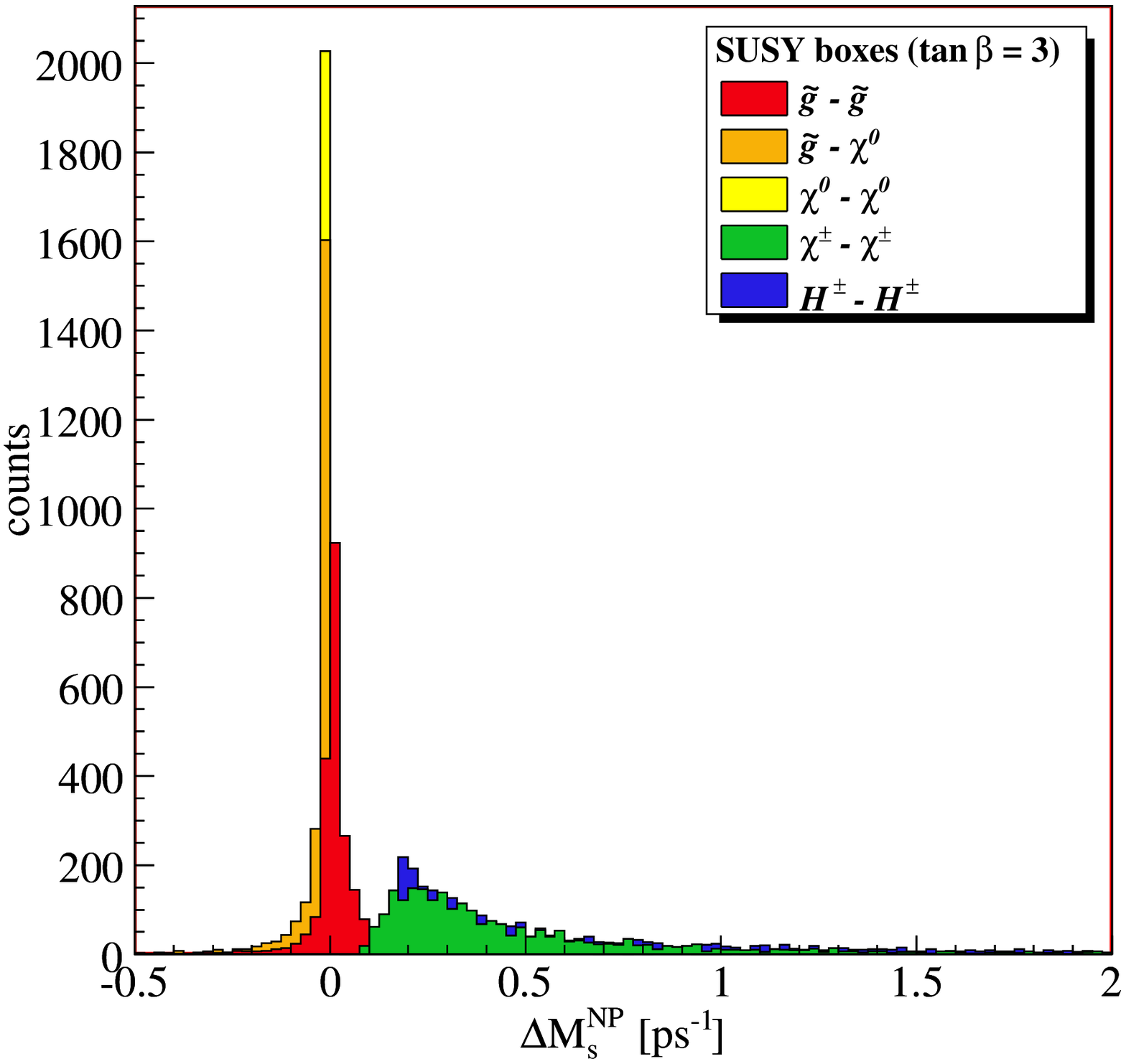}
\includegraphics[scale=0.35]{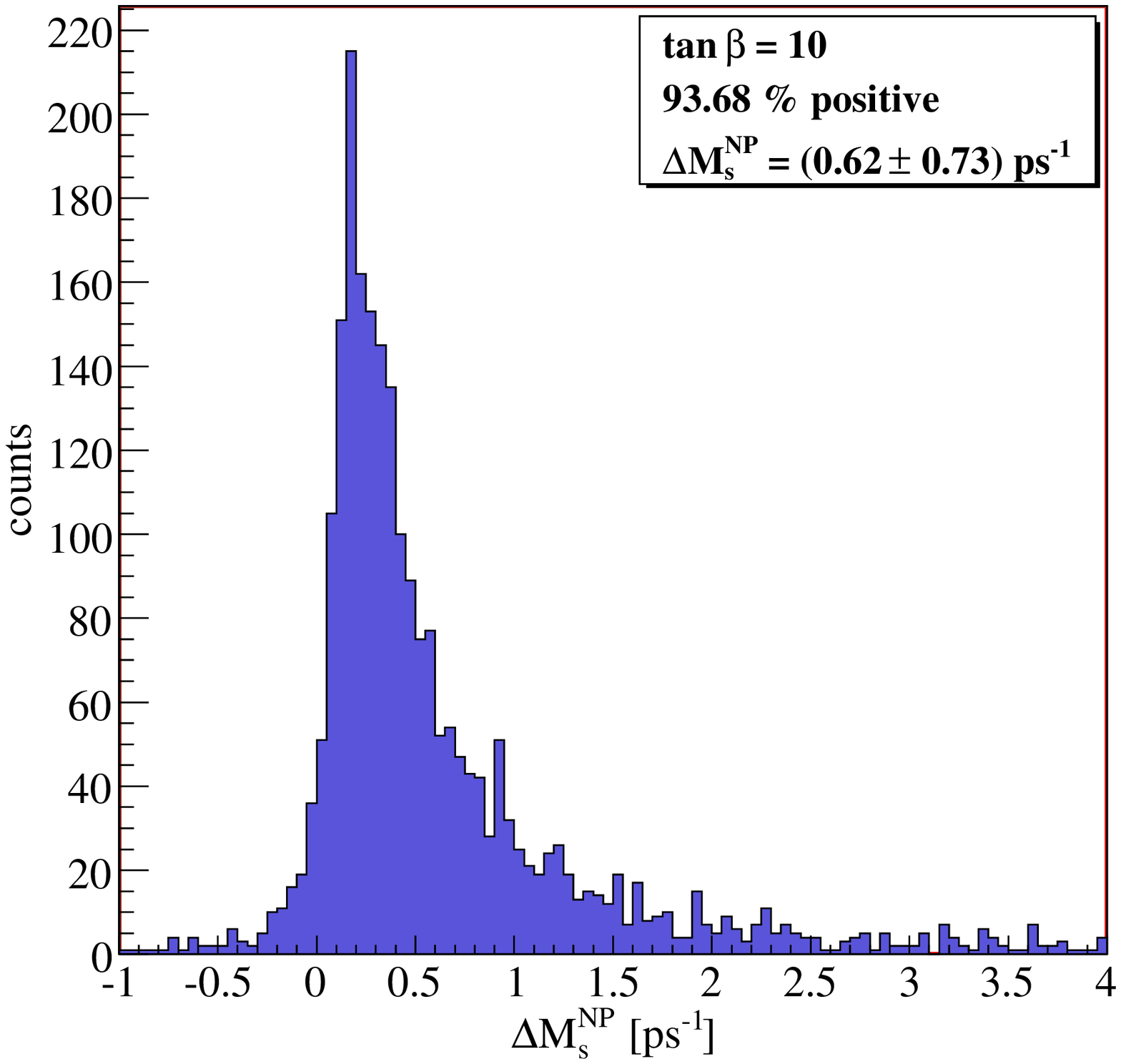}
\includegraphics[scale=0.35]{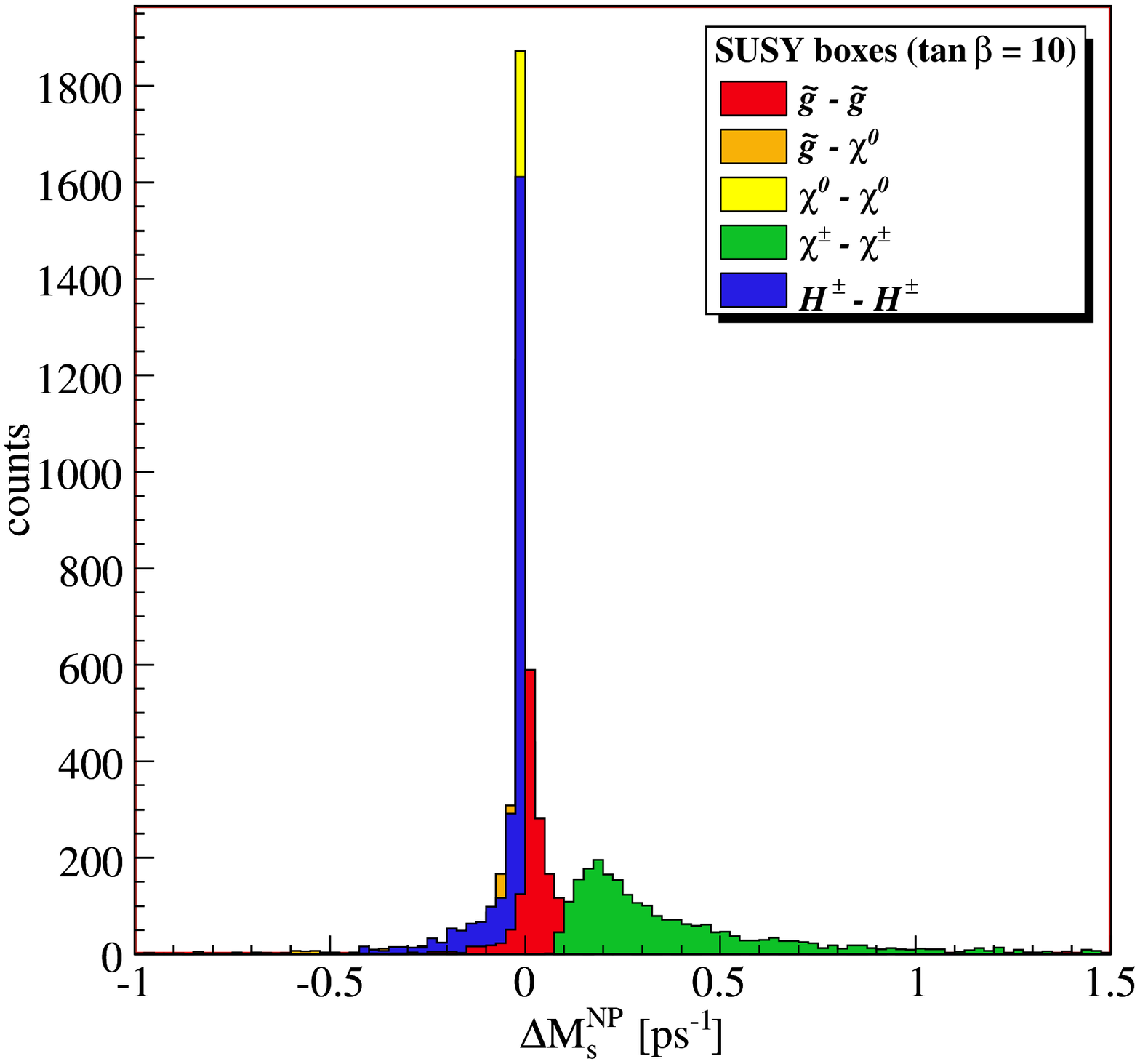}
\end{center}
\caption{\small\sl Distributions of values for $\D M_s^{\rm NP}$ in the MFV MSSM: sum of
the contributions (left panels) and separate SUSY contributions (right panels).
The distributions result from scanning the MFV parameters $a_i$, $b_i$ as well as the
SUSY mass scales (see text for details). First and second row plots refer to $\tan \be =
3$ and $10$, respectively.}
\label{fig:runtb}
\end{figure}

The results of such global runs are reported in Fig. \ref{fig:runtb} for the cases of
$\tan \be = 3$ and $10$.
As the Figure shows, for $\tan \be = 3$ the positiveness of the sum of results is
confirmed by the global run as well.\footnote{We mention that a further verification was
carried out by varying the `B-parameters' for the effective operator matrix elements
within the ranges allowed in \cite{Bparams-lattice-DB}. Neither the 
positiveness of the sum of the results, nor the interplay beween charginos and gluinos, 
resulting in the above regimes (A),...,(D), are touched at all by variations of the
B-parameters.} The right panel of the first row also shows that, if $M_{H^{\pm}_1}$ 
is assumed to be flatly distributed in the range $M_{H^{\pm}_1} \in [90,1000]$ GeV, then 
Higgs contributions tend to give a distribution completely analogous to that for charginos 
(the latter hides almost entirely the former in the plot).

The second row of Fig. \ref{fig:runtb} reports then the same distributions, when 
$\tan \be = 10$. As one can see, contributions from charginos, gluinos, neutralinos and
mixed gluino-neutralino are left qualitatively unchanged by the variation in $\tan \be$.
On the other hand, Higgs contributions are very small and negative, as expected
also from the lowermost curve in Fig. \ref{fig:Higgs}. However, this fact has quite a
small impact on the sum of all the results, which remains chiefly positive (see left
panel). We mention that a further increase in $\tan \be$ to $15$ confirms a similar
picture: all contributions but for Higgs boxes build up an almost totally positive
distribution. The negativity of Higgs boxes, however, starts to matter, and the sum of 
all contributions is positive in roughly $86$ \% of the counts. 

\subsubsection*{$\bullet$ Changing the ranges for MFV parameters}

\noindent A second set of runs was devised to check for possible variations in our findings, when
the MFV parameters defining the expansions (\ref{softMFV}) are varied in ranges different
from those specified in eq. (\ref{aibi-values}). In the latter, the choice of $1$ as the
upper bound is dictated by various considerations. First, on `aesthetical' grounds, 
such bound should be realistic, if the MFV expansion of the soft terms as functions of 
the Yukawa couplings is a `natural' one. The interpretation is that $a_i$ and $b_i$ `couplings' of O(1) 
cause the new flavour violating effects originated by the soft terms to be at the same time 
{\em CKM-like} -- \ie generated by the SM Yukawa couplings, according to the very MFV 
paradigm -- and {\em natural} \cite{hall-randall-MFV,MFV}.
Second, on more technical grounds, enlarging too much the range of variation for the $a_i$
and $b_i$ may lead to a non-efficient exploration of the full parameter space allowed.
However, on this latter point, we have verified that the shape of the distributions
remains stable already after a few hundreds random values for the $a_i, b_i$.

In this second MonteCarlo study, we kept SUSY masses fixed to scenarios, as described 
in Section \ref{sec:strategy}, and increased the upper bounds for the MFV parameters from 
$1$ to $5$. The main effect of increasing the ranges of variability is to `smoothen' the
difference between chargino and gluino distributions in the various scenarios and
consequently to make regimes (A),..., (D) more similar to one another. However, the main
features of each of them, as well as the positiveness of the sum of contributions, are 
left unchanged.

On the whole, a choice of the MFV parameters according to the ranges (\ref{aibi-values})
-- mainly dictated by `naturalness' -- allows a better understanding of the interplay
between chargino and gluino contributions in the different scenarios. It would be interesting 
to adopt a top-down approach to the determination of the MFV parameters $a_i, b_i$, by 
considering \eg SUSY Grand Unified Theories (GUT) which at low energy reproduce the soft SUSY 
breaking structure of the MSSM. If such SUSY GUTs can be constructed to be minimal flavour 
violating \cite{GCIW}, then the $a_i, b_i$ at the EW breaking scale can be {\em fitted} by 
introducing the corresponding constraints (\ref{softMFV}) between soft terms and Yukawa
couplings. 
The latter are both {\em predicted} within such models, in terms of the initial conditions
at the GUT scale and of the running.

\bigskip

We conclude this section by stressing that the feature of positiveness for the sum of SUSY 
contributions to $\D M_s$ in the MFV MSSM is a precise signature of low 
$\tan \be \simle 10$. It is interesting to address the question whether a similar signature, 
but with sign reversed, applies to the MFV MSSM in the large $\tan \be$ limit. This regime 
requires however consideration of an additional set of contributions, represented by the 
double neutral Higgs penguins \cite{BCRS-NP,isidori-retico,BCRS-PL,BCRSbig,FGH}. This goes 
beyond the scope of the present paper. For an comprehensive related study in the context of
grand unification, see Ref. \cite{LPV}.

\newpage
\section{Various considerations on MFV} \label{sec:limits-MFV}

The concrete application of the MFV limit to a calculable NP model, like the MSSM, gives us 
now the opportunity to emphasize the main differences between the model independent approach 
of \cite{MFV} and the former, phenomenological definition of MFV by \cite{BurasMFV}.
We will also try and understand in which, among our studied scenarios for the SUSY scales,
MFV contributions to meson oscillations are dominated by $\mc Q_1$, the operator already
present within the SM. This case is referred to as `constrained' MFV (CMFV) \cite{BBGT}
and its study will allow us to understand how natural CMFV is within the MSSM.

\subsection{MFV: definition \cite{MFV} versus \cite{BurasMFV}}

According to the definition of MFV by \cite{MFV}, all the flavour and CP violation at the
EW scale is generated by the Yukawa couplings present in the SM. As we stressed several times
above, this does not mean that new sources of flavour violation should be set to zero, but 
instead that they should be taken as functions of the SM Yukawa couplings. The functional dependence 
is in turn fixed by their formal transformation properties under the flavour group and gives 
rise to expansions like (\ref{softMFV}).

On the other hand, the phenomenological definition of MFV by \cite{BurasMFV} does not
start from the SM Yukawa couplings, but from the CKM matrix. In this approach, a theory is considered 
minimal flavour violating, when only interactions which display {\em explicit} proportionality 
to the CKM matrix are active. In this context, new sources of flavour violation are simply
not considered, since they have a priori ``nothing to do'' with the CKM matrix. 

However, in the more general approach by \cite{MFV}, the CKM matrix becomes a {\em by-product} 
of the SM Yukawa couplings. Then the possibility of treating the Yukawa couplings as spurion fields, allows to 
classify in terms of them any {\em new} source of flavour violation, and this is why the 
definition by \cite{MFV} has a completely general applicability. As a matter of fact, it
also provides a framework for systematic routes to beyond-MFV (see \cite{NMFV-FM}).

In the general context of \cite{MFV}, let us finally establish a contact with definition
\cite{BurasMFV}. Since in the approach \cite{BurasMFV} new flavour violating structures are 
taken as {\em unrelated} to the SM Yukawa couplings, it is clear that one can reach this 
definition by setting the MFV parameters $b_i \to 0$, in expansions (\ref{softMFV}).
In this respect, an interesting MFV study where effects beyond the less general 
framework \cite{BurasMFV} are not visible has been reported in Ref. \cite{IMPST}.
Specifically, in this case squark flavor mixing effects proportional to Yukawa couplings are 
negligible (in particular there are no gluino contributions) and the $b_i$ coefficients
are consistently set to zero.

\subsection{Deviations from `constrained' MFV}

In this Section, we finally study -- in the context of the $\D B = 2$ Hamiltonian -- 
the special case in which MFV MSSM contributions are dominated by the operator $\mc Q_1$,
the one already present in the SM. This case corresponds to `constrained' MFV (CMFV)
\cite{BBGT}\footnote{For the sake of clarity, we mention that the definition of MFV
adopted in \cite{BBGT} complies with \cite{BurasMFV}.}. We would like to stress that CMFV 
is a phenomenologically relevant limit: in fact, if MFV is motivated by the observation 
that experiments do not require new sources of flavour violation besides the SM Yukawa
couplings, 
the constrained version of it is analogously justified by the fact that there is no compelling 
experimental evidence for contributions of effective operators other than the 
SM ones. 

The latter statement is actually strictly true when one supposes that new effective 
operators be ``strongly coupled'', \ie multiplied by an overall `coupling' factor taken to be $1$
(or $-1$) \cite{LEP-paradox,MFV}. Bounds on new operator contributions are considerably
alleviated, or removed altogether, when such operators are weakly coupled, as is typically
the case for SUSY contributions to the observables considered in the bounds.

The limit of CMFV is a useful benchmark case to consider, both because of the
generic argument on new operators mentioned above and because it is a very well satisfied
feature of chargino contributions, which represent an important and sometimes dominating 
ingredient in the various mass regimes of MFV.
In the present Section, we start again from our calculated MFV MSSM contributions 
to meson oscillations. We then compare -- in the different mass scenarios considered -- 
the part of the contributions which bears proportionality to $\mc Q_1$ with the rest 
of the contributions, due to operators besides the SM one. This exercise will allow us 
to understand how `natural' is CMFV within the MSSM.

As a first step, we can define the ratio $R_{\rm CMFV}$ as
\beqn
R_{\rm CMFV} \equiv \frac{|\D M_s^{\rm NP}(C_1 \to 0)|}{\D M_s^{\rm NP}({\rm only}~C_1)}~,
\label{R}
\eeqn
where $C_i$ are the Wilson coefficients of the $\D B = 2$ Hamiltonian (see eq.
(\ref{Ci-MSSM})) in the MFV MSSM. The ratio $R_{\rm CMFV}$ provides 
a good `quantitative' understanding of the relative importance of contributions from 
non-SM operators with respect to $\mc Q_1$, in the $B_s$ case.

\begin{figure}[t]
\begin{center}
\includegraphics[scale=0.35]{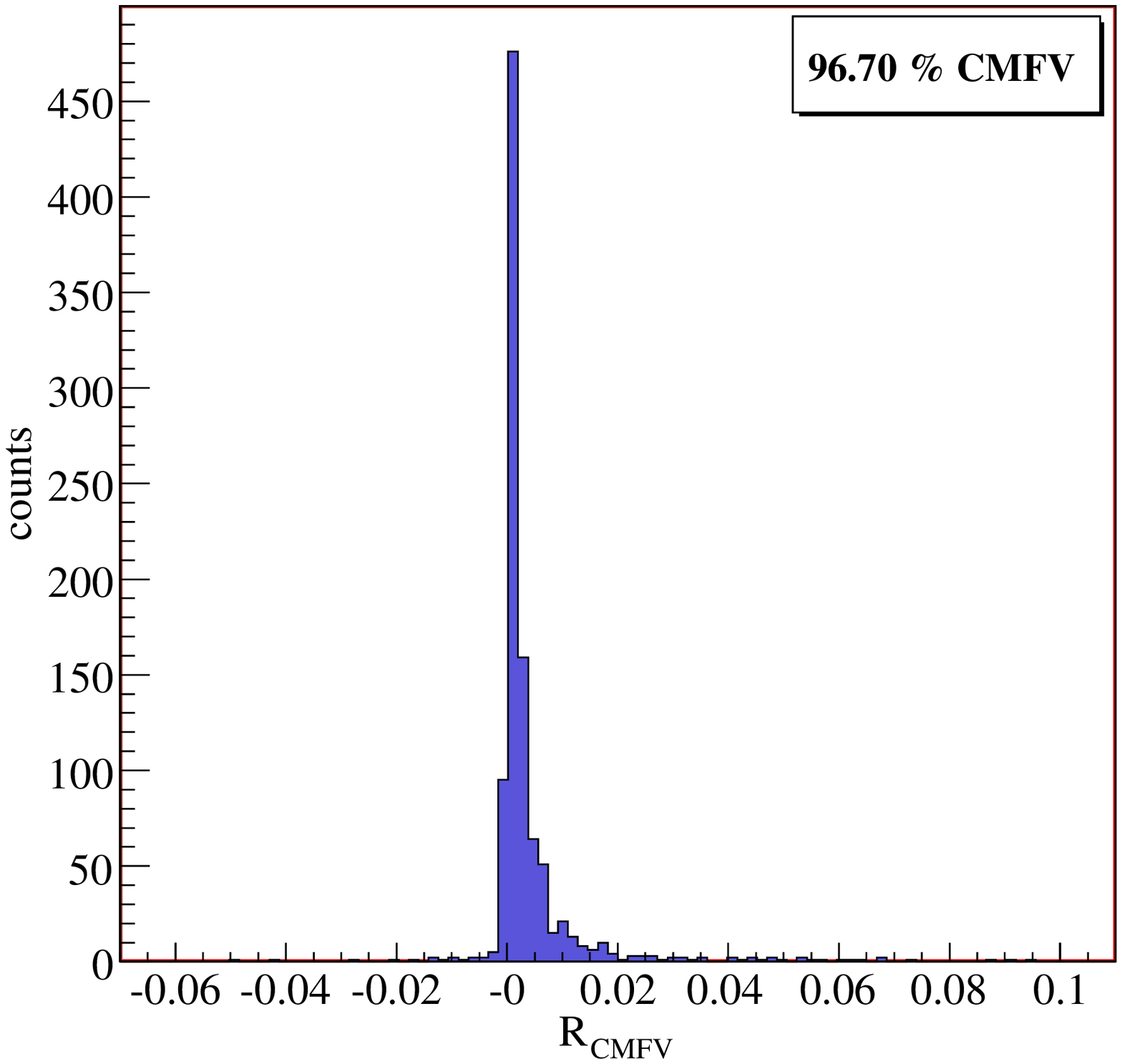}
\includegraphics[scale=0.35]{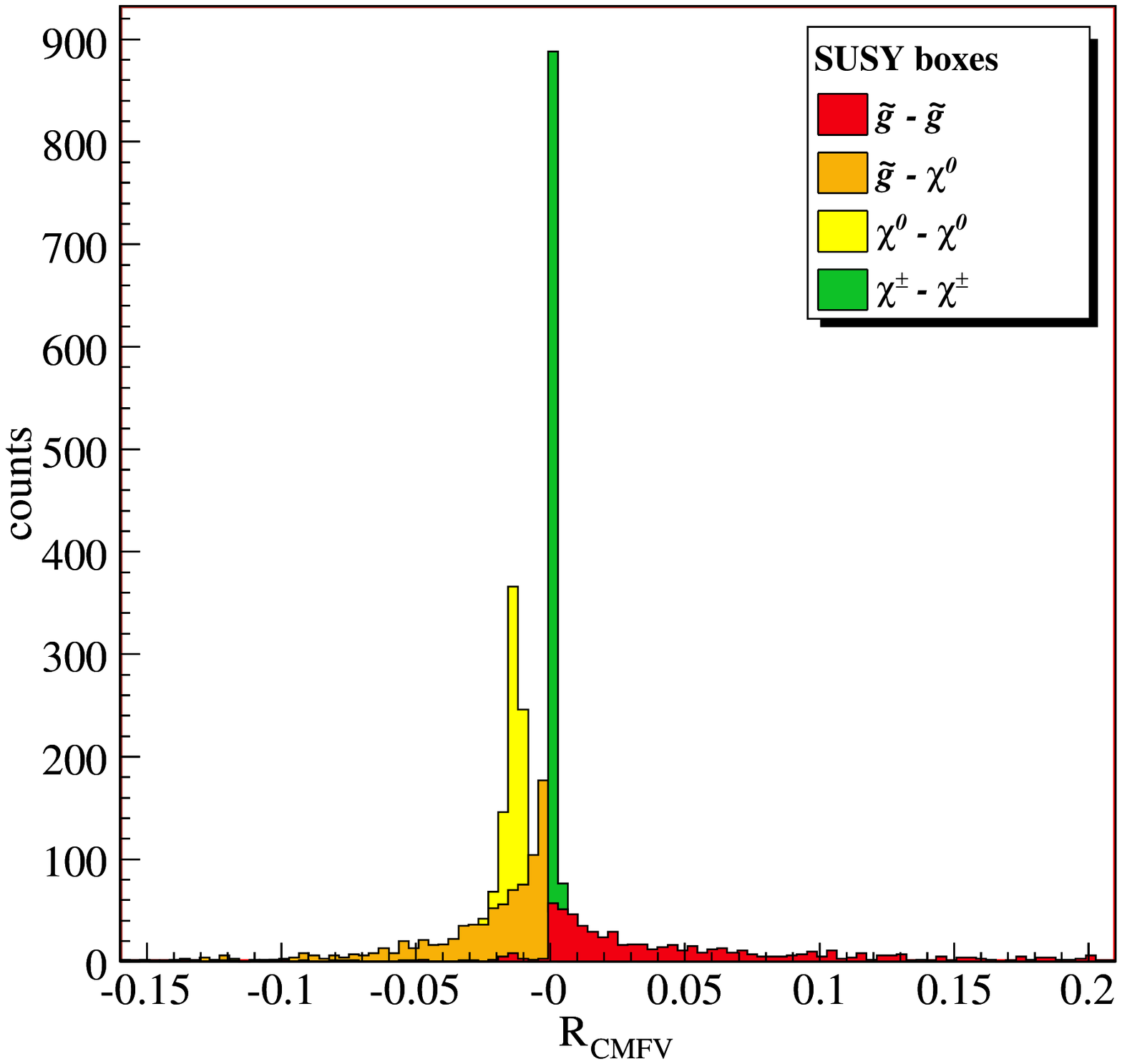}
\end{center}
\caption{\small\sl Ratio (\ref{R}) of the contributions to $\D M_s^{\rm NP}$ from operators other
than $\mc Q_1$ to that from $\mc Q_1$ alone, for the set of SUSY scales chosen in Fig.
\ref{fig:sum-mu200}, which is representative of mass regime (A). Left panel shows the sum of
all contributions but for charged Higgs boxes, right panel reports the ratio for the
separate contributions.}
\label{fig:CiC1-A}
\end{figure}

\begin{figure}[t]
\begin{center}
\includegraphics[scale=0.35]{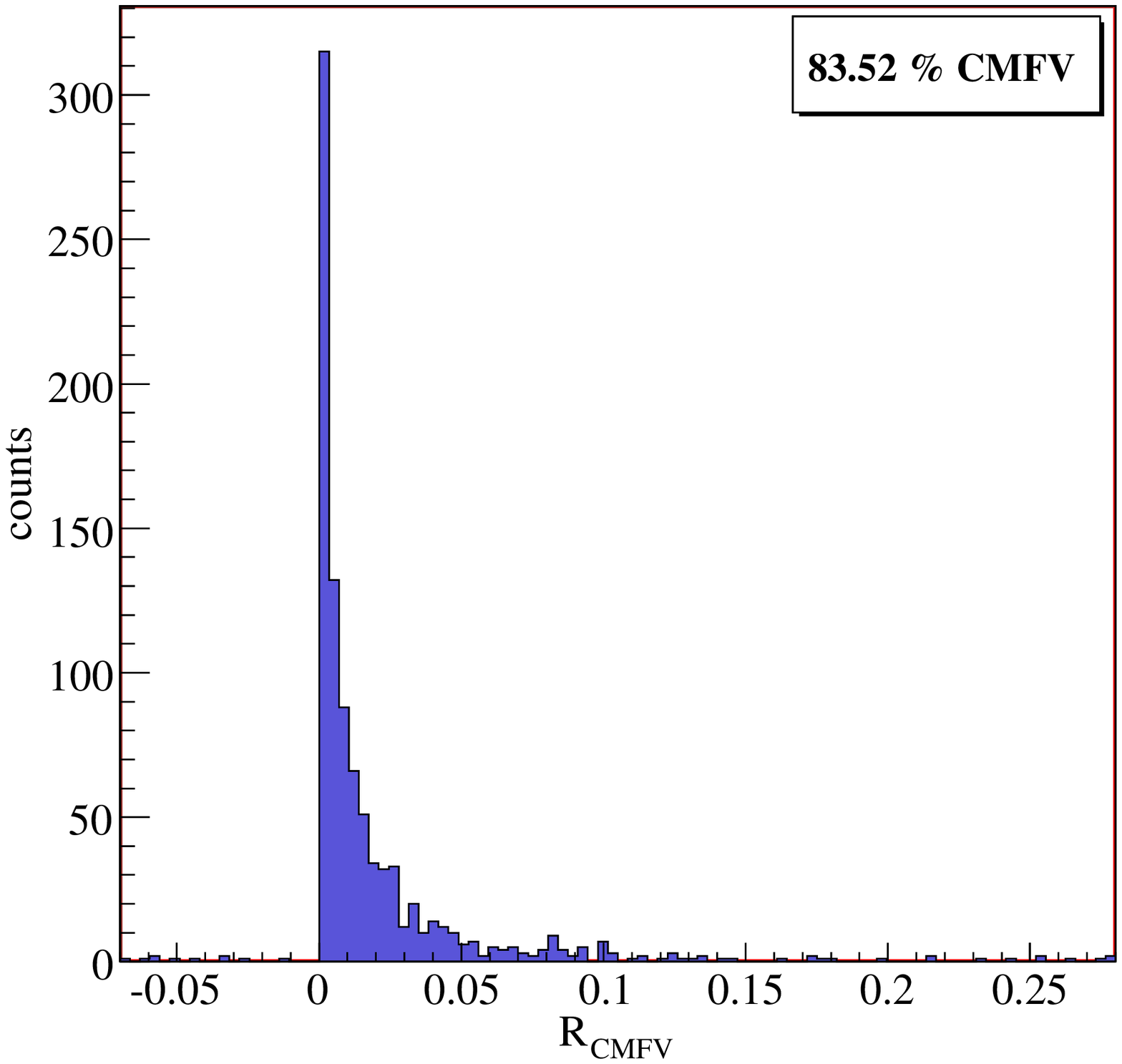}
\includegraphics[scale=0.35]{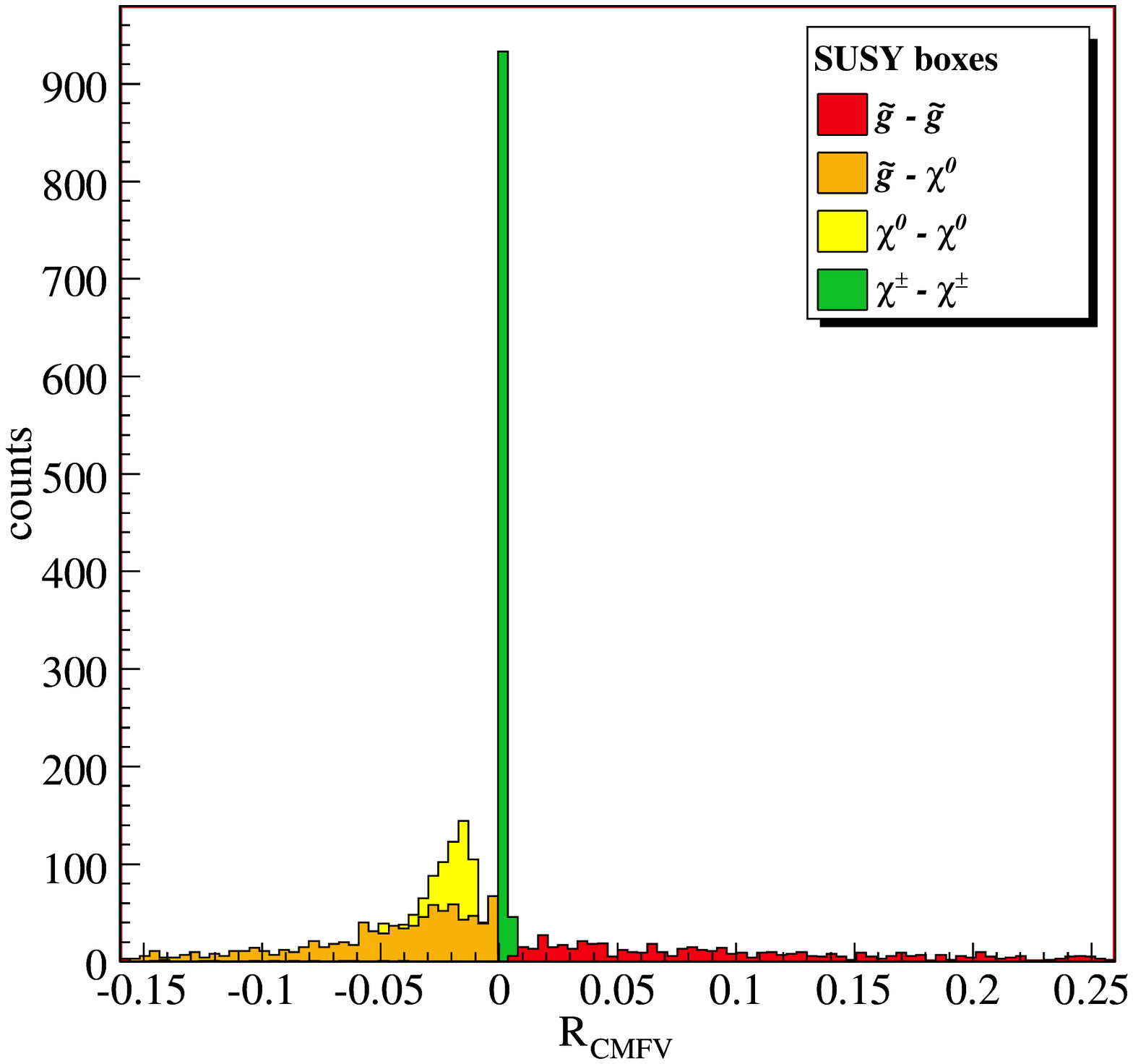}
\end{center}
\caption{\small\sl Same as Fig. \ref{fig:CiC1-A}, but for the set of SUSY scales, chosen in
this case as in Fig. \ref{fig:sum-mu500}, which is representative of mass regime (B).}
\label{fig:CiC1-B}
\end{figure}

\begin{figure}[t]
\begin{center}
\includegraphics[scale=0.35]{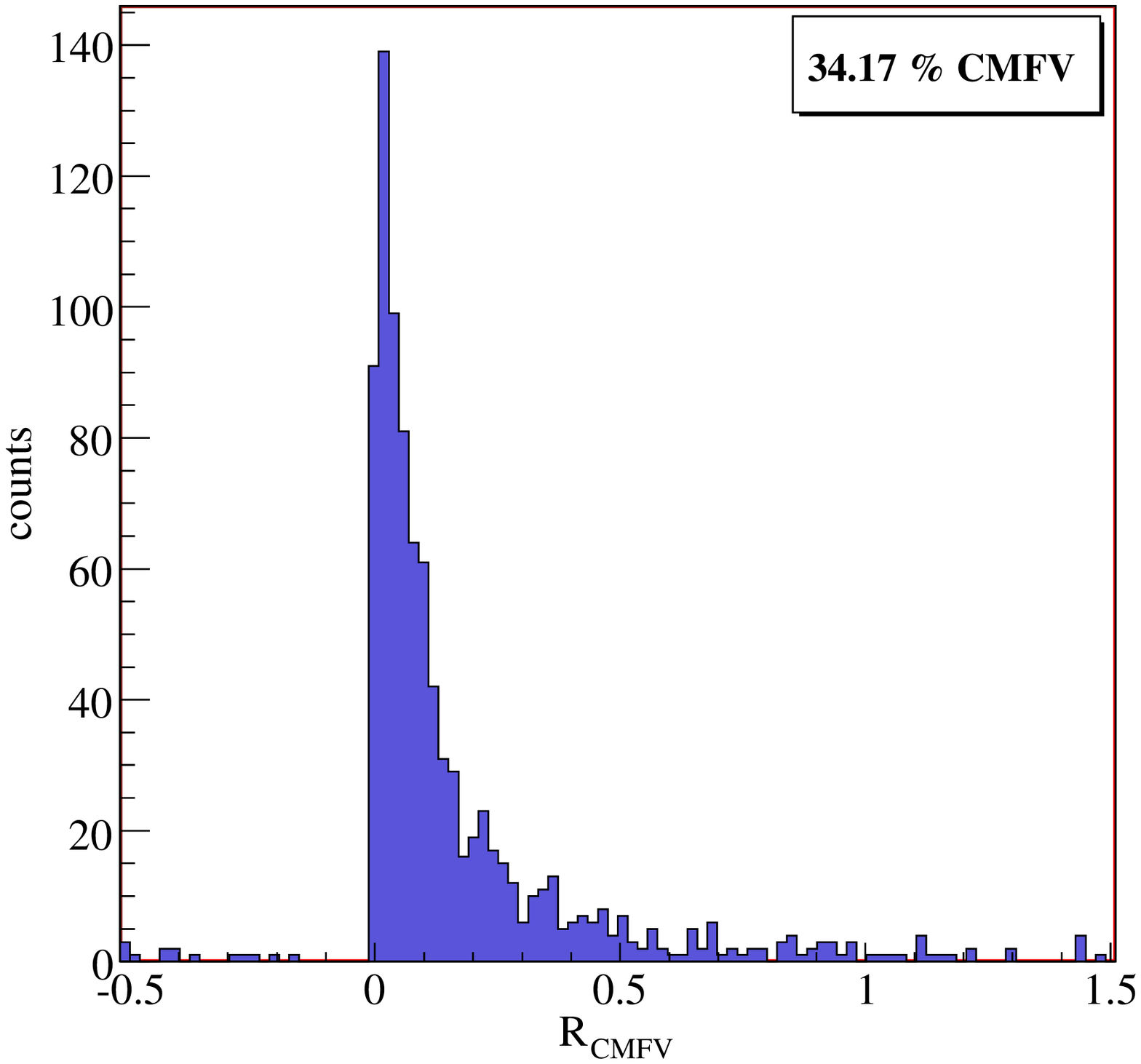}
\includegraphics[scale=0.35]{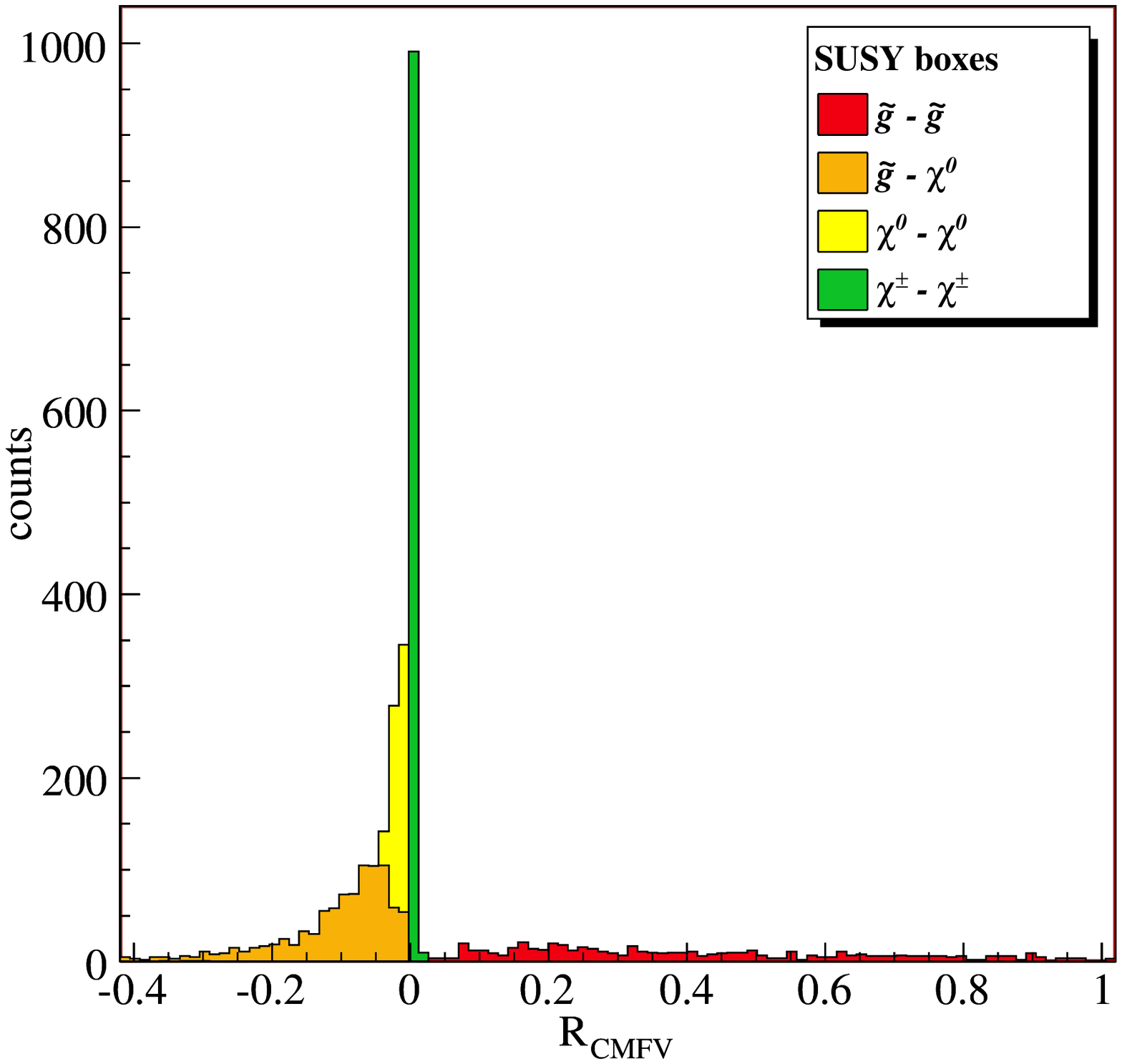}
\end{center}
\caption{\small\sl Same as Fig. \ref{fig:CiC1-A}, but for the set of SUSY scales, chosen in
this case as in Fig. \ref{fig:sum-mu1000}, which is representative of mass regime (C).}
\label{fig:CiC1-C}
\end{figure}

\begin{figure}[t]
\begin{center}
\includegraphics[scale=0.35]{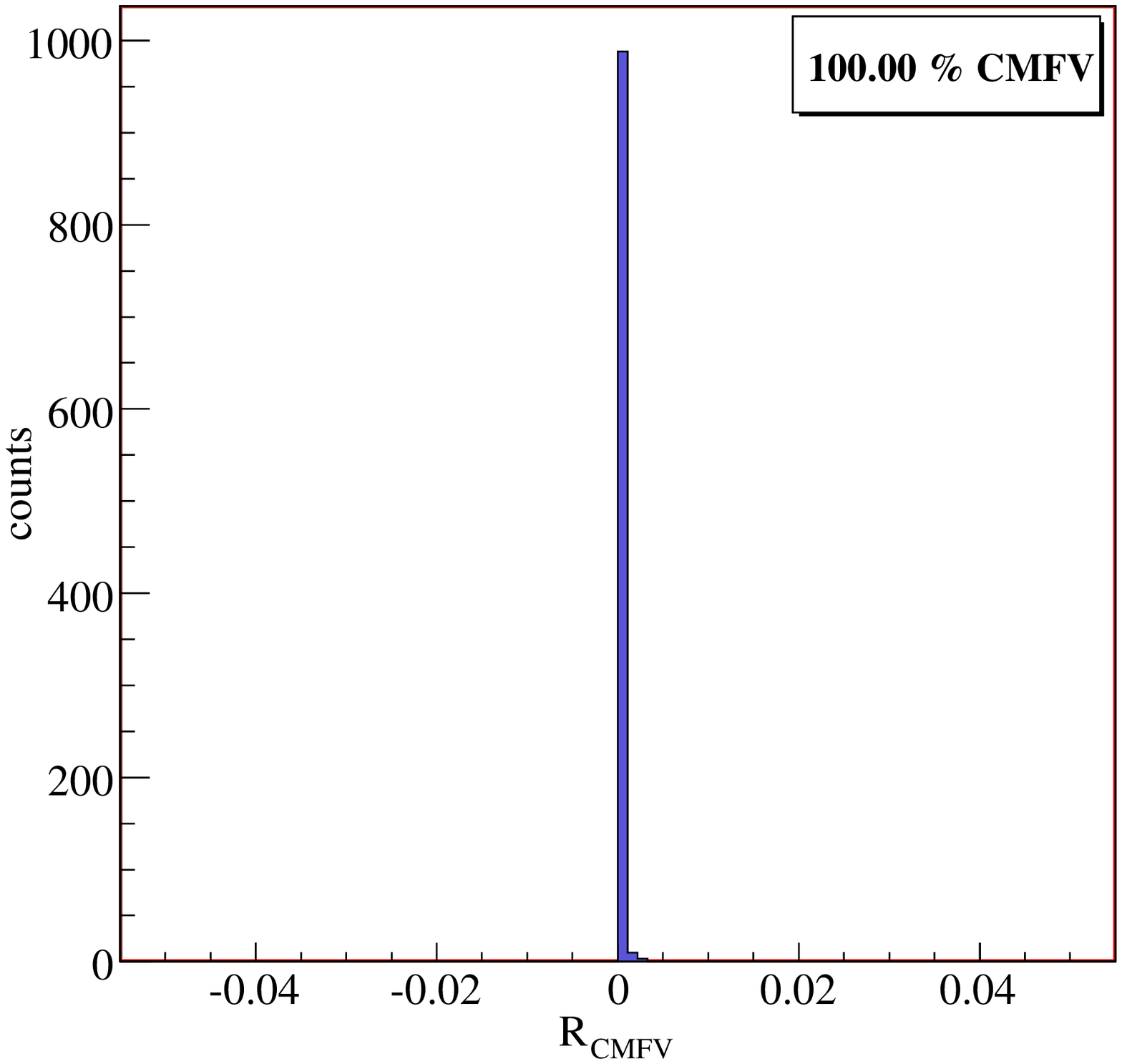}
\includegraphics[scale=0.35]{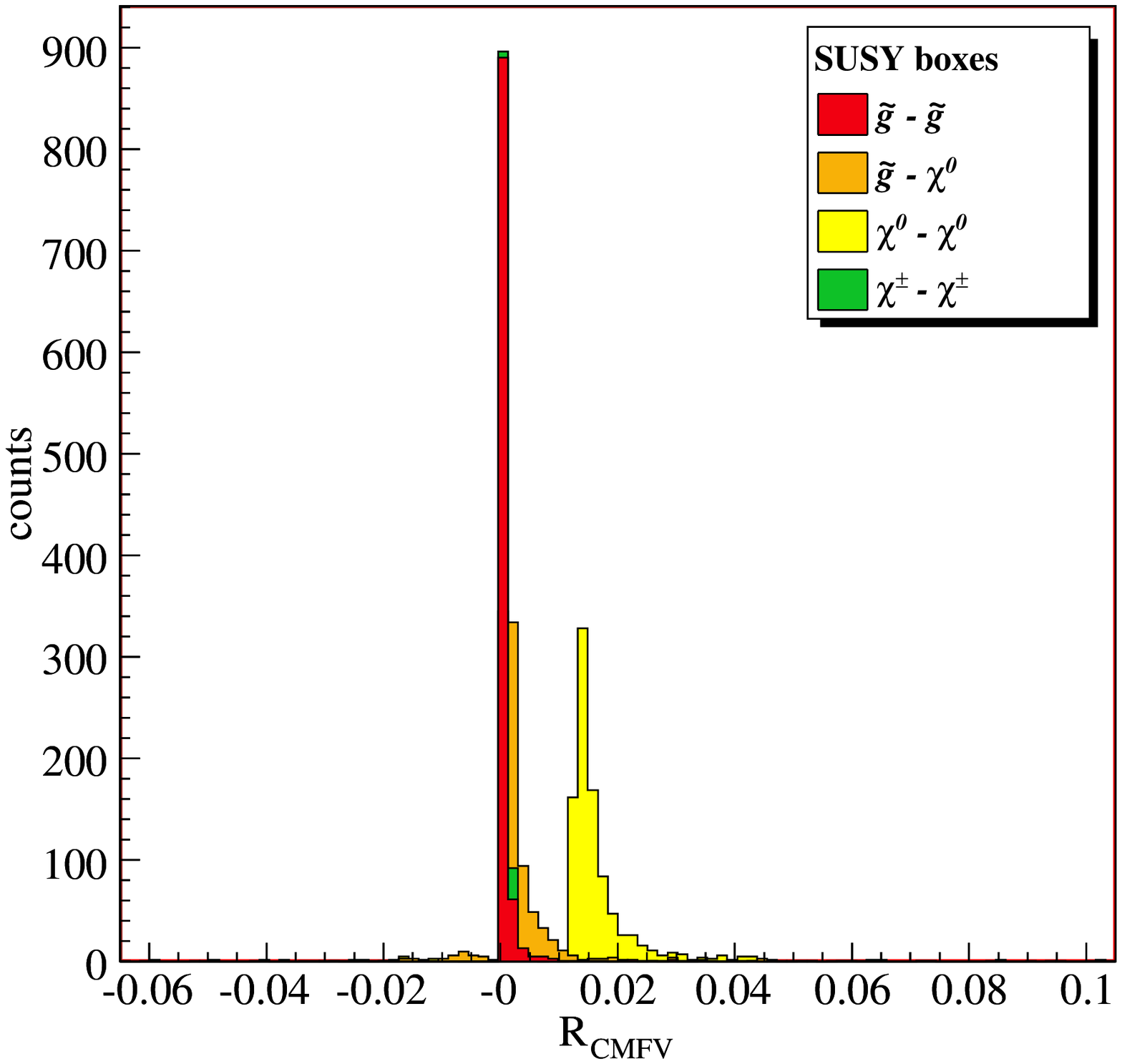}
\end{center}
\caption{\small\sl Same as Fig. \ref{fig:CiC1-A}, but for the set of SUSY scales, chosen in
this case as in Fig. \ref{fig:sum-bigMsq}, which is representative of mass regime (D).}
\label{fig:CiC1-D}
\end{figure}

Within every of the mass scenarios considered, the ratio $R_{\rm CMFV}$ is then a
distribution of values depending on the MFV parameters $a_i, b_i$.
In Figs. \ref{fig:CiC1-A}-\ref{fig:CiC1-D} we report such distribution in the four
representative regimes studied in Section \ref{sec:results}. In particular, left panels
display $R_{\rm CMFV}$ for the sum of all contributions, leaving aside, again, those from
charged Higgs, on which we will comment separately. The percentage reported on top of
the plots gives the number of counts satisfying $|R_{\rm CMFV}| \le 0.05$. This bound can
be considered -- in our MFV MSSM case -- a quantitative `definition' of CMFV. According to
it, a given estimate of $\heff$ is taken to be CMFV, if contributions from operators other 
than $\mc Q_1$ do not exceed $5$ \% of the contribution from $\mc Q_1$. We warn the reader 
that this upper bound (and the corresponding definition associated with it) has of course a large 
degree of arbitrariness: we chose $5$ \% in view of the still large hadronic uncertainties 
associated with the estimate of the $\< \mc Q_i \>$.

As one can see from the left panels of Figs. \ref{fig:CiC1-A}-\ref{fig:CiC1-D}, mass
regimes (A) and (D) satisfy well CMFV. Mass regimes (B) and in particular (C), on the
other hand, do not.\footnote{We mention that the case displayed in Fig. \ref{fig:CiC1-B} is
the one with the highest CMFV percentage among the studied scenarios belonging to mass
regime (B).} Further insight can be gained by looking at the right panels of the
same Figures, where $R_{\rm CMFV}$ is displayed for the single contributions (in this case
the ratio (\ref{R}) was calculated by restricting separately to $C_i^{\tilde g \tilde g}$
for gluino-gluino boxes, and so on for the other contributions). One immediately
recognizes that regime (A) is $\mc Q_1$-dominated, since chargino contributions are the
most important ones. Gluinos are not CMFV, but they are unimportant. 
In regime (D), instead, charginos and gluinos are of the same size, but both $\mc
Q_1$-dominated. Finally, in regimes (B) and especially (C), CMFV does not apply: in fact, while
charginos are always important and $\mc Q_1$-dominated, gluinos are not negligible and not
$\mc Q_1$-dominated. As a matter of fact, $R_{\rm CMFV}$ for gluinos is in these cases
very spread, typical values being in the range $20 \div 40$. We mention that such large 
values are not shown in the plots of Figs. \ref{fig:CiC1-B} and \ref{fig:CiC1-C}.

Let us then address how the above picture is modified when adding contributions from charged 
Higgses. The $R_{\rm CMFV}$ for the latter, in the case of $\tan \be = 3$, ranges between 
$0.035$ and $0.015$, for $M_{H_1^\pm} = 90$ GeV and $1000$ GeV respectively. However, 
contributions from scalar operators steadily increase in importance when increasing $\tan \be$: 
for $\tan \be = 4$ Higgs contributions are not CMFV if $M_{H_1^\pm} \simle 500$ GeV and,
already from $\tan \be = 5$, in the full range of masses considered for $M_{H_1^\pm}$.
Then, whether the total sum is CMFV or not, it depends on how large the contribution from 
Higgses is, with respect to the other contributions, \ie on the Higgs mass chosen (see
Fig. \ref{fig:Higgs}).

We remark at this point that an increase in importance of scalar contributions affects, 
in general, $\D M_s$ and $\D M_d$ in a different way. However, in the case of Higgs
boxes, we found that the ratio of $\D M_d/\D M_s$ has a sensitivity with respect to 
variations of $M_{H_1^\pm}$ below $1$ \%, when $\tan \be$ is strictly small. Variations
start to be visible only from $\tan \be \simge 10$, when they become $> 3$ \%. This point
was already stated at the end of Section \ref{sec:MC}, when discussing the case of 
$\D M_d$.

Concerning contributions other than charged Higgses, we note that large deviations from 
$\mc Q_1$-dominated MFV apply when $\mu$ is not small -- regimes (B) and (C) --. However,
such deviations turn out not to affect $\D M_s$ and $\D M_d$ in a sensibly different way,
so that their impact is not visible in the ratio $\D M_d/\D M_s$. The reason was, again, 
already explained at the end of Section \ref{sec:MC}.

However, the above discussion allows us to place a warning message on the
conventional wisdom that, for low $\tan \be$, MFV in the $\D F = 2$ Hamiltonian is 
$\mc Q_1$-dominated. This is clearly not true when Higgs and gluino contributions are included, 
and when $\mu$ is not small in magnitude. In this case, the approximation $Y_d \to 0$, on the 
ground that $\tan \be$ is small, turns out not to work.

\subsection[A lower bound on $\D M_{s,d}$ from CMFV]{\boldmath A lower bound on $\D M_{s,d}$ from CMFV}

Recently, it has been shown analytically in \cite{BB} that, within the CMFV models, 
the lower bounds on $\D M_{d,s}$ are simply given by $(\D M_{d,s})_{\rm SM}$. The proof
holds, assuming general exchange of charged gauge bosons, Goldstone bosons and physical 
scalars in boxes together with Dirac fermions, and under phenomenologically realistic assumptions 
on their masses. On the other hand, two possible exceptions to the argument were found to arise 
in the presence of Majorana fermions and of $U(1)$ neutral gauge bosons in box diagrams, which could 
individually bring $\D M_{d,s}$ below $(\D M_{d,s})_{\rm SM}$. Within the
model-independent approach of \cite{BB}, it seems impossible to exclude CMFV models involving 
{\em negative} contributions to $\D M_{d,s}$, caused by Majorana fermions and/or $U(1)$ neutral 
gauge bosons, that are not cancelled by the remaining {\em positive} contributions from the 
other possible particle exchanges.

Within our study, we found in the previous Section that only mass regimes (A) and (D) are 
$\mc Q_1$-dominated, \ie CMFV. In particular, the above mentioned exception, of a 
CMFV case in which Majorana fermion contributions are not negligible, arises in regime (D) 
(see Fig. \ref{fig:sum-bigMsq}). However in this case, Majorana contributions, represented by the
gluino boxes, are also positive, so that, at least for low $\tan \be$, the lower bounds of 
\cite{BB} still hold.
Finally, regime (A) exactly corresponds to the assumptions made in the proof of \cite{BB},
since Majorana contributions are negligible. In this case we numerically confirm the
positivity of the NP contributions.

We find actually remarkable that, within the MFV MSSM at low $\tan \be$, the condition 
$\D M_{s,d} > (\D M_{s,d})_{\rm SM}$ holds even when new operators give non-negligible
contributions.

\section{MFV-Unitarity Triangle} \label{sec:MFV-UUT}

In this Section, we would like to emphasize that the so-called Universal Unitarity Triangle 
(UUT) \cite{BurasMFV} is valid only within the CMFV models, and not within the general class of
MFV models, like the MSSM considered here.

Indeed, the usual construction of the UUT is based on the value of the angle $\be$,
measured by means of the $S_{\psi K_S}$ asymmetry, and on the value of the side $R_t$,
obtained from the ratio $\D M_d/\D M_s$, that within CMFV is independent of any new
physics contributions.

As already stressed in \cite{BCRSbig,BBGT}, in the context of the MSSM at large $\tan
\be$, the presence of new operator contributions to $\D M_{s}$ and $\D M_{d}$ generally
modifies the relation between $R_t$ and $\D M_d / \D M_s$, so that it now reads \cite{BCRSbig,BBGT}
\beqn
R_t = 0.913 \left[ \frac{\xi}{1.23} \right] \sqrt{\frac{17.8/{\rm ps}}{\D M_s}}
\sqrt{\frac{\D M_d}{0.507/{\rm ps}}} \sqrt{R_{sd}}~, ~~~~R_{sd} = \frac{1+f_s}{1+f_d}~,
\label{Rt}
\eeqn
with $\D M_q = (\D M_q)_{\rm SM}(1+f_q)$ and $\xi$ being a known non-perturbative
parameter, $\xi = 1.23(6)$.

\begin{figure}[t]
\begin{center}
\includegraphics[scale=1.0]{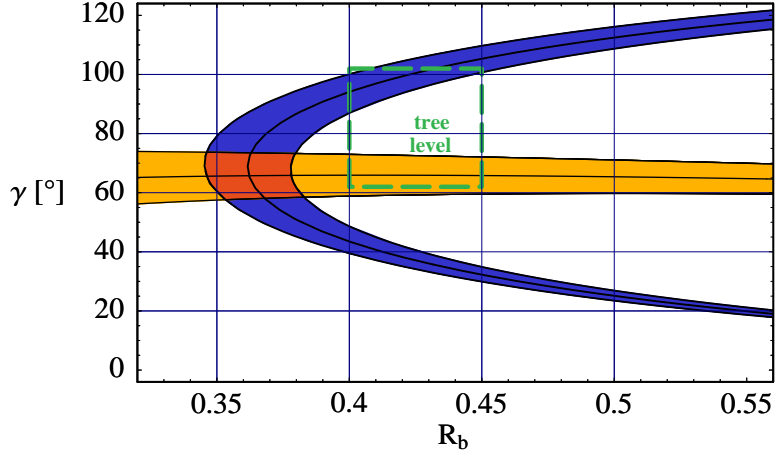}
\end{center}
\caption{\small\sl The blue area represents the correlation between $R_b$ and $\ga$, 
valid in MFV. The orange area corresponds to the region with $R_t$ fixed to the value 
in eq. (\ref{Rt}). CMFV occurs in the intersection between the two areas, displayed in
red. Finally, the green dashed box shows the $1\sigma$-allowed range for $R_b$ and $\ga$,
as measured from tree-level decays.}
\label{fig:Rb-vs-gamma}
\end{figure}
Basing on the results of the previous sections, formula (\ref{Rt}) applies to the MSSM at 
low  $\tan \be$ as well, with $R_{sd} \ne 1$, due to the presence of new operators that 
differently affect $\D M_s$ and $\D M_d$. This can already be seen by adding to the SM 
expression for the mass
differences, the contributions from the charged Higgs boxes, which do not depend on the
MFV parameters $a_i, b_i$. As we showed in Section \ref{sec:limits-MFV}, when addressing
deviations from CMFV, the ratio $\D M_d / \D M_s$ is subject to variations at the percent 
level, depending on the value chosen for $M_{H_1^\pm}$. Similar effects occur in
principle when considering the other contributions as well, but in this case the 
variation is completely hidden by the lack of knowledge of the values for the MFV parameters 
$a_i, b_i$ and of course of the SUSY scales. The departure from CMFV is more pronounced
in the MFV MSSM for large $\tan \be$, but such analysis is beyond the scope of the present 
paper.

The above argument clearly demonstrates that the MFV MSSM, even at low $\tan \be$,
does not belong to the class of CMFV models, and consequently, the UUT obtained from 
$R_t$ and $\be_{\psi K_S}$ is not generically valid in MFV.

This discussion suggests that the unitarity triangle valid for all MFV models is not the
UUT of \cite{BurasMFV} -- and analyzed in detail by \cite{UTfit} -- but a triangle
constructed from the angle $\be_{\psi K_S}$ and the new-physics-independent value of 
$|V_{ub}|$ or $\ga$ from tree-level decays.

In this context, the known value $\be_{\psi K_S} = (21.2 \pm 1.0)^\circ$ establishes a
correlation between the side $R_b$,
\beqn
R_b = \left( 1- \frac{\la^2}{2} \right) \frac{1}{\la} 
\left| \frac{V_{ub}}{V_{cb}} \right|~,
\eeqn
of the unitarity triangle and the angle $\ga$, that is valid for all models with MFV. 
Such correlation is represented in Fig. \ref{fig:Rb-vs-gamma} as a blue area, under the 
assumption $\be = \be_{\psi K_S} = (21.2 \pm 1.0)^\circ$. The orange area in the Figure
displays instead the region characterized by $R_t$ fixed to the value in eq. (\ref{Rt}),
with $R_{sd} = 1$ and $\xi = 1.23(6)$. The intersection between the two areas 
-- displayed in red -- is then the one allowed to CMFV. Fig. \ref{fig:Rb-vs-gamma} also
shows the $1 \sigma$-allowed range from tree-level decays, namely $62^\circ \le \ga \le
102^\circ$ and $0.40 \le R_b \le 0.45$, as a green dashed box. The latter overlaps with
the higher branch of the MFV area but not with the CMFV one. This indicates, on the one
hand, somehow a `tension' \cite{BBGT} between the tree-level determination of $R_b$ and 
the one favoured by CMFV. On the other hand, the overlap between the green box and the
blue area suggests that this tension disappears within MFV, provided $\ga > 80^\circ$.

It will be interesting to monitor with the help of Fig. \ref{fig:Rb-vs-gamma} the progress
in the determination of $R_b$ and of the angle $\ga$ from tree-level decays, and to verify
whether the correlation in question is satisfied by the data.

Equivalently, with precise values of $\ga$ and $R_b$, that are used for the construction
of the reference unitarity triangle, it will be possible to find out whether the CMFV UUT
or the MFV-UT is chosen by nature, or instead if one has to introduce non-MFV interactions 
to fit the data. This would occur if the experimental point $(\ga, R_b)$ lies outside the 
blue area in Fig. \ref{fig:Rb-vs-gamma}.

\section{Conclusions and Outlook} \label{sec:conclusions}

In this paper, we have applied the effective field theory definition of 
Minimal Flavour Violation (MFV) to the MSSM and explicitly shown how, by this definition, 
the new sources of flavour and CP violation present in the MSSM become functions of the SM 
Yukawa couplings. We have subsequently applied the MFV limit to the MSSM $\Delta B = 2$ Hamiltonian 
at low $\tan \be$.

Our findings can be summarized in the following main messages:
\begin{enumerate}

\item In the MFV limit of the MSSM, the soft breaking terms become functions of the
SM Yukawa couplings. This non-trivial functional dependence causes flavour violation due to 
soft terms to be not zero in MFV, but instead `CKM-like'. The constraints imposed on the
soft terms by the MFV assumption lead to a significant increase in the predictive power of
the model.

\item The supersymmetric corrections to $\D M_{s,d}$ at low $\tan \be$ are found to be
always {\em positive} with respect to the SM formula. This feature is due to an
interesting interplay between the chargino and gluino box diagram contributions.

\item Even at low $\tan \be$, the MSSM does not in general belong to the class of models 
with CMFV, in contrast to the statements made in the literature. The presence of gluino 
box diagram contributions necessarily brings in new operators beyond the
$(V-A)\otimes(V-A)$ one, whose importance depends on the mass regime chosen.

\item The side $R_t$, used in the determination of the Universal Unitarity Triangle, is
actually not a good constant in MFV. Within the MFV MSSM at low $\tan \be$, we have found 
variations that can reach the percent level, depending again on the mass regime chosen, and 
on the value of $\tan \be$. To resolve such variations, one however needs a lattice 
determination for $\xi$ with an error at most half the present value. In the case of the
MFV MSSM at large $\tan \be$ the situation is in this respect `easier', since larger
deviations from the CMFV value of $R_t$ are more likely.

\item A unitarity triangle valid in all MFV models (MFV-UT) can be constructed using only 
$|V_{ub}|$ or $\ga$, from tree-level decays, and the angle $\be$, extracted from 
$S_{\psi K_S}$. In particular, with the measured value of $\be_{\psi K_S}$, MFV implies a 
testable correlation between $|V_{ub}|$ and $\ga$. With the present high value of $|V_{ub}|$, 
MFV favours $\ga > 80^\circ$. The LHC program on the measurement of $\gamma$ is then of utmost 
importance to cleanly consolidate -- or disprove -- the MFV paradigm.

\end{enumerate}
\medskip

\section*{Acknowledgments}
A.J.B. would like to thank Gino Isidori for his critical remarks in connection with the
bound \cite{BB}, that strengthened the motivation to carry out this study. D.G is indebted 
to Gino Isidori for insightful conversations on the topic of MFV, that were precious for
framing the subject of this work.
The authors thank also Monika Blanke for a careful reading of the manuscript and useful
comments. Thanks are also due to Uli Haisch and Federico Mescia for useful comments. 
This work has been supported in part by the Cluster of Excellence ``Origin and 
Structure of the Universe'' and by the German Bundesministerium f{\"u}r Bildung und Forschung
under contract 05HT6WOA.

\appendix
\renewcommand{\theequation}{\thesection.\arabic{equation}}




\def \mutype#1{m_{u_{#1}}}
\def \mdtype#1{m_{d_{#1}}}

\def \mU#1{M_{U_{#1}}}
\def \mD#1{M_{D_{#1}}}

\def \mw{M_{W}}
\def \mz{M_{Z}}
\def \mhpm#1{M_{H^{\pm}_{#1}}}

\def \mcha#1{M_{\chi_{#1}}}
\def \mneu#1{M_{\chi^{0}_{#1}}}
\def \mg{M_{\tilde g}}




\appendix

\section{\boldmath Wilson coefficients for the $\D B = 2$ effective 
Hamiltonian in the MSSM}\label{app:CW}

\subsection{Contributions}

Below we list the non-vanishing new physics contributions to the Wilson coefficients for
$B_s$ mixing, eq. (\ref{Ci-MSSM}). The case of $B_d$ mixing is obtained by the replacement 
$2 \to 1$ (or $5 \to 4$ where applicable) in the external quark indices.
In the following expressions, it is always understood that internal indices are summed over 
their respective ranges, \textit{i.e.} $I,J,K = 1,...,3$ 
for quarks, $i,j = 1,...,6$ for squarks, $a,b = 1,2$ for charginos and $a,b = 1,...,4$ 
for neutralinos.

\subsubsection*{Charged Higgs contributions}
%
\begin{multline}
	C_1^{H^+H^+} = \frac{g_2^4}{16\pi^2} K_{I3} K_{J3} K_{I2}^{*} K_{J2}^{*} \frac{\mutype{I}^2 \mutype{J}^2}{8 \mw^4} \left\{ 2 \mw^2  D_0(m^2_{u_I},m^2_{u_J},\mhpm{1}^2,\mw^2) \cot^2\beta \right. \\
	- \left. D_2(m^2_{u_I},m^2_{u_J},\mhpm{1}^2,\mhpm{1}^2)\cot^4\beta - 2 D_2(m^2_{u_I},m_{u_J}^2,\mhpm{1}^2,M^2_{W})\cot^2\beta \right\} ~,\nonumber
\end{multline}
\begin{multline}
	\tilde C_1^{H^+H^+} = -\frac{g_2^4}{16\pi^2} K_{I3} K_{J3} K_{I2}^{*} K_{J2}^{*}  \frac{m_s^2 m_b^2}{8 \mw^4} \\
	\times \Big\{ D_2(m^2_{u_I},m^2_{u_J},\mhpm{1}^2,\mhpm{1}^2) \tan^4\beta + 2 D_2(m^2_{u_I},m^2_{u_J},\mhpm{1}^2,\mw^2) \tan^2\beta \Big\} ~,\nonumber
\end{multline}
\begin{multline}
	C_2^{H^+H^+} = - \frac{g_2^4}{16\pi^2} K_{I3} K_{J3} K_{I2}^{*} K_{J2}^{*}  \frac{m_s^2 \mutype{I}^2 \mutype{J}^2}{8 \mw^4} \\ 
	\times \Big\{ D_0(m_{u_I}^2,m_{u_J}^2,\mhpm{1}^2,\mhpm{1}^2) - 2 D_0(m_{u_I}^2,m_{u_J}^2,\mhpm{1}^2,\mw^2) \Big\} ~, \nonumber
\end{multline}
\begin{multline}
	\tilde C_2^{H^+H^+} = - \frac{g_2^4}{16\pi^2} K_{I3} K_{J3} K_{I2}^{*} K_{J2}^{*}  \frac{m_b^2 \mutype{I}^2 \mutype{J}^2}{8 \mw^4} \\ 
	\times \Big\{ D_0(m_{u_I}^2,m_{u_J}^2,\mhpm{1}^2,\mhpm{1}^2) - 2 D_0(m_{u_I}^2,m_{u_J}^2,\mhpm{1}^2,\mw^2) \Big\} ~,\nonumber
\end{multline}
%
%
\begin{multline}
	C_4^{H^+H^+} = \frac{g_2^4}{16\pi^2} K_{I3} K_{J3} K_{I2}^{*} K_{J2}^{*} \left\{ \frac{m_s m_b}{\mw^2} D_2(m^2_{u_I},m^2_{u_J},\mhpm{1}^2,M^2_{W}) \tan^2\beta \right. \\
	-  \frac{m_s m_b m^2_{u_I} m^2_{u_J}}{4 \mw^4}  \Big( D_0(m^2_{u_I},m^2_{u_J},\mhpm{1}^2,\mhpm{1}^2)\\
	+  D_0(m^2_{u_I},m^2_{u_J},\mhpm{1}^2,\mw^2) \left( \tan^2\beta + \cot^2\beta \right)\Big) \bigg\} ~,\nonumber
\end{multline}
\begin{multline}
	C_5^{H^+H^+} = \frac{g_2^4}{16\pi^2} K_{I3} K_{J3} K_{I2}^{*} K_{J2}^{*}  \frac{m_s m_b \mutype{I}^2}{2 \mw^4} \\
	\times \Big\{ D_2(m^2_{u_I},m^2_{u_J},\mhpm{1}^2,\mhpm{1}^2) - 2 D_2(m^2_{u_I},m^2_{u_J},\mhpm{1}^2,\mw^2) \Big\} ~.\nonumber
\end{multline}
The above coefficients $C_i^{H^+ H^+}$ and $\tilde C_i^{H^+ H^+}$ arise from the sum of 
$H^+_1 - H^+_1$, $H^+_1 - H^+_2$ and $H^+_1 - W^+$ boxes (with $H^+_2$ the charged Goldstone boson, 
entering calculations away from the unitary gauge).
\subsubsection*{Chargino contributions}
%
\begin{equation}
	C_1^{\chi^+\chi^+}  =  -\frac{1}{32\pi^2} D_2(\mU{i}^2,\mU{j}^2,\mcha{a}^2,\mcha{b}^2) \left[ V_{\chi Ud}^L \right]_{ai3} \left[ V_{\chi Ud}^L \right]_{bj3} \left[ V_{\chi Ud}^L \right]^{*}_{aj2} \left[ V_{\chi Ud}^L \right]^{*}_{bi2}~, \nonumber
\end{equation}
%
%
\begin{equation}
	C_3^{\chi^+\chi^+} = - \frac{\mcha{a} \mcha{b}}{32\pi^2} D_0(\mU{i}^2,\mU{j}^2,\mcha{a}^2,\mcha{b}^2) \left[ V_{\chi Ud}^L \right]_{ai3}\left[ V_{\chi Ud}^L \right]_{bj3}\left[ V_{\chi Ud}^R \right]^{*}_{aj2} \left[ V_{\chi Ud}^R \right]^{*}_{bi2}~, \nonumber
\end{equation}
%
%
\begin{equation}
	C_4^{\chi^+\chi^+} = \frac{1}{8\pi^2} D_2(\mU{i}^2,\mU{j}^2,\mcha{a}^2,\mcha{b}^2)\left[ V_{\chi Ud}^R \right]_{ai3} \left[ V_{\chi Ud}^L \right]_{bj3} \left[ V_{\chi Ud}^R \right]^{*}_{aj2} \left[ V_{\chi Ud}^L \right]^{*}_{bi2}~,\nonumber
\end{equation}
\begin{equation}
	C_5^{\chi^+\chi^+} = -\frac{\mcha{a} \mcha{b}}{16\pi^2} D_0(\mU{i}^2,\mU{j}^2,\mcha{a}^2,\mcha{b}^2)\left[ V_{\chi Ud}^R \right]_{ai3} \left[ V_{\chi Ud}^L \right]_{bj3} \left[ V_{\chi Ud}^L \right]^{*}_{aj2} \left[ V_{\chi Ud}^R \right]^{*}_{bi2}~.\nonumber
\end{equation}
The coefficients $\tilde C_1^{\chi^+\chi^+}$ and $\tilde C_3^{\chi^+\chi^+}$ can be obtained 
from $C_1^{\chi^+\chi^+}$ and $C_3^{\chi^+\chi^+}$ respectively by the replacement of
left- and right-handed chargino couplings $V^L \leftrightarrow V^R$.

\subsubsection*{Neutralino contributions}
%
\begin{multline}
	C_1^{\chi^0\chi^0} = - \frac{1}{32\pi^2} D_2(\mD{i}^2,\mD{j}^2,\mneu{a}^2,\mneu{b}^2) \left[ V_{\chi Dd}^L \right]^{*}_{bi2}\left[ V_{\chi Dd}^L \right]^{*}_{aj2}\left[ V_{\chi Dd}^L \right]_{bj3} \left[ V_{\chi Dd}^L \right]_{ai3}  \\
	- \frac{\mneu{a} \mneu{b}}{64\pi^2} D_0(\mD{i}^2,\mD{j}^2,\mneu{a}^2,\mneu{b}^2) \left[ V_{\chi Dd}^L \right]^{*}_{bi2}\left[ V_{\chi Dd}^L \right]^{*}_{bj2}\left[ V_{\chi Dd}^L \right]_{aj3} \left[ V_{\chi Dd}^L \right]_{ai3}~,\nonumber
\end{multline}
%
%
\begin{equation}
	C_2^{\chi^0\chi^0} = \frac{\mneu{a} \mneu{b}}{32\pi^2} D_0(\mD{i}^2,\mD{j}^2,\mneu{a}^2,\mneu{b}^2)  \left[ V_{\chi Dd}^L \right]_{ai3} \left[ V_{\chi Dd}^R \right]^{*}_{bi2} \left[ V_{\chi Dd}^L \right]_{aj3}\left[ V_{\chi Dd}^R \right]^{*}_{bj2}~,\nonumber
\end{equation}
%
%
\begin{multline}
	C_3^{\chi^0\chi^0}  =  -\frac{\mneu{a} \mneu{b}}{32\pi^2} D_0(\mD{i}^2,\mD{j}^2,\mneu{a}^2,\mneu{b}^2) \left[ V_{\chi Dd}^L \right]_{ai3} \left[ V_{\chi Dd}^R \right]^{*}_{bi2} \\
	\times \left\{\left[ V_{\chi Dd}^L \right]_{bj3}\left[ V_{\chi Dd}^R \right]^{*}_{aj2} - \left[ V_{\chi Dd}^L \right]_{aj3}\left[ V_{\chi Dd}^R \right]^{*}_{bj2}\right\}~,\nonumber
\end{multline}
%
%
\begin{multline}
	C_4^{\chi^0\chi^0} = \frac{1}{8\pi^2} D_2(\mD{i}^2,\mD{j}^2,\mneu{a}^2,\mneu{b}^2) \left[ V_{\chi Dd}^R \right]_{ai3} \left[ V_{\chi Dd}^L \right]^{*}_{bi2} \\
	\times \left\{ \left[ V_{\chi Dd}^L \right]_{bj3} \left[ V_{\chi Dd}^R \right]^{*}_{aj2} + \left[ V_{\chi Dd}^L \right]_{aj3} \left[ V_{\chi Dd}^R \right]^{*}_{bj2} \right\}~,\nonumber
\end{multline}
\begin{multline}
	C_5^{\chi^0\chi^0} = - \frac{1}{8\pi^2} D_2(\mD{i}^2,\mD{j}^2,\mneu{a}^2,\mneu{b}^2) \left[ V_{\chi Dd}^L \right]^{*}_{bi2}\left[ V_{\chi Dd}^R \right]^{*}_{bj2}\left[ V_{\chi Dd}^L \right]_{ai3} \left[ V_{\chi Dd}^R \right]_{aj3}  \\
	- \frac{\mneu{a} \mneu{b}}{16\pi^2} D_0(\mD{i}^2,\mD{j}^2,\mneu{a}^2,\mneu{b}^2) \left[ V_{\chi Dd}^L \right]^{*}_{aj2}\left[ V_{\chi Dd}^R \right]^{*}_{bi2}\left[ V_{\chi Dd}^L \right]_{bj3} \left[ V_{\chi Dd}^R \right]_{ai3}~.\nonumber
\end{multline}
Interchanging left- with right-handed neutralino couplings in $C_{1,2,3}^{\chi^0\chi^0}$ yields the expressions for the Wilson coefficients $\tilde C_{1,2,3}^{\chi^0\chi^0}$.

\subsubsection*{Mixed neutralino and gluino contributions}
%
\begin{multline}
	C_1^{\tilde g \chi 0} = -\frac{g_3^2}{16\pi^2} \frac{2}{3} D_2(\mD{i}^2,\mD{j}^2,\mneu{a}^2,\mg) (Z_D)_{3j}(Z_D)^{*}_{2i}\left[ V_{\chi Dd}^L \right]_{ai3} \left[ V_{\chi Dd}^L \right]_{aj2}^{*} \\
	- \frac{g_3^2}{16\pi^2} \frac{\mneu{a} \mg}{6}  D_0(\mD{i}^2,\mD{j}^2,\mneu{a}^2,\mg) \\
	\times \left\{ (Z_D)_{3i}(Z_D)_{3j} \left[ V_{\chi Dd}^L \right]_{ai2}^{*} \left[ V_{\chi Dd}^L \right]_{aj2}^{*} + (Z_D)_{2i}^{*}(Z_D)_{2j}^{*} \left[ V_{\chi Dd}^L \right]_{ai3} \left[ V_{\chi Dd}^L \right]_{aj3} \right\}~,\nonumber
\end{multline}
%
%
\begin{multline}
	C_2^{\tilde g \chi 0} = \frac{g_3^2}{16\pi^2} \frac{\mneu{a} \mg}{3} D_0(\mD{i}^2,\mD{j}^2,\mneu{a}^2,\mg) \left\{ 3 (Z_D)_{3j}(Z_D)^{*}_{5i}\left[ V_{\chi Dd}^L \right]_{ai3} \left[ V_{\chi Dd}^R \right]_{aj2}^{*} \right. \\
	\left. + (Z_D)_{3i}(Z_D)_{3j} \left[ V_{\chi Dd}^R \right]_{ai2}^{*} \left[ V_{\chi Dd}^R \right]_{aj2}^{*} + (Z_D)_{5i}^{*}(Z_D)_{5j}^{*} \left[ V_{\chi Dd}^L \right]_{ai3} \left[ V_{\chi Dd}^L \right]_{aj3} \right\}~,\nonumber
\end{multline}
%
%
\begin{multline}
	C_3^{\tilde g \chi 0} =  -\frac{g_3^2}{16\pi^2} \frac{\mneu{a} \mg}{3} D_0(\mD{i}^2,\mD{j}^2,\mneu{a}^2,\mg) \left\{ (Z_D)_{3j}(Z_D)^{*}_{5i}\left[ V_{\chi Dd}^L \right]_{ai3} \left[ V_{\chi Dd}^R \right]_{aj2}^{*}  \right. \\
	\left. - (Z_D)_{3i}(Z_D)_{3j} \left[ V_{\chi Dd}^R \right]_{ai2}^{*} \left[ V_{\chi Dd}^R \right]_{aj2}^{*} -(Z_D)_{5i}^{*}(Z_D)_{5j}^{*} \left[ V_{\chi Dd}^L \right]_{ai3} \left[ V_{\chi Dd}^L \right]_{aj3} \right\}~,\nonumber
\end{multline}
%
%
\begin{multline}
	C_4^{\tilde g \chi 0} = -\frac{g_3^2}{16\pi^2} \frac{2}{3} D_2(\mD{i}^2,\mD{j}^2,\mneu{a}^2,\mg)   \\
	\times \left\{ (Z_D)_{3j}(Z_D)^{*}_{2i} \left[ V_{\chi Dd}^R \right]_{ai3} \left[ V_{\chi Dd}^R \right]_{aj2}^{*} + (Z_D)_{6j}(Z_D)_{5i}^{*} \left[ V_{\chi Dd}^L \right]_{ai3} \left[ V_{\chi Dd}^L \right]_{aj2}^{*} \right.  \\
	- (Z_D)_{3j}(Z_D)_{6i} \left[ V_{\chi Dd}^L \right]^{*}_{ai2} \left[ V_{\chi Dd}^R \right]_{aj2}^{*} - (Z_D)_{2j}^{*}(Z_D)_{5i}^{*} \left[ V_{\chi Dd}^L \right]_{ai3} \left[ V_{\chi Dd}^R \right]_{aj3}  \\
	\left. - 3 (Z_D)_{6i}(Z_D)_{3j} \left[ V_{\chi Dd}^L \right]_{aj2}^{*} \left[ V_{\chi Dd}^R \right]_{ai2}^{*} - 3 (Z_D)_{5i}^{*}(Z_D)_{2j}^{*} \left[ V_{\chi Dd}^L \right]_{aj3} \left[ V_{\chi Dd}^R \right]_{ai3} \right\}  \\
	+ \frac{g_3^2}{16\pi^2} \mneu{a} \mg D_0(\mD{i}^2,\mD{j}^2,\mneu{a}^2,\mg)   \\
	\times \left\{ (Z_D)_{3j}(Z_D)_{5i}^{*} \left[ V_{\chi Dd}^R \right]_{ai3} \left[ V_{\chi Dd}^L \right]_{aj2}^{*} + (Z_D)_{6j}(Z_D)_{2i}^{*} \left[ V_{\chi Dd}^L \right]_{ai3} \left[ V_{\chi Dd}^R \right]_{aj2}^{*} \right\} ~,\nonumber
\end{multline}
\begin{multline}
	C_5^{\tilde g \chi 0} =  \frac{g_3^2}{16\pi^2} \frac{2}{3} D_2(\mD{i}^2,\mD{j}^2,\mneu{a}^2,\mg)  \\
	\times \left\{ 3 (Z_D)_{3j}(Z_D)^{*}_{2i} \left[ V_{\chi Dd}^R \right]_{ai3} \left[ V_{\chi Dd}^R \right]_{aj2}^{*} + 3 (Z_D)_{6j}(Z_D)_{5i}^{*} \left[ V_{\chi Dd}^L \right]_{ai3} \left[ V_{\chi Dd}^L \right]_{aj2}^{*} \right.  \\
	- 3 (Z_D)_{3j}(Z_D)_{6i} \left[ V_{\chi Dd}^L \right]^{*}_{ai2} \left[ V_{\chi Dd}^R \right]_{aj2}^{*} - 3 (Z_D)_{2j}^{*}(Z_D)_{5i}^{*} \left[ V_{\chi Dd}^L \right]_{ai3} \left[ V_{\chi Dd}^R \right]_{aj3}  \\
	\left. - (Z_D)_{6i}(Z_D)_{3j} \left[ V_{\chi Dd}^L \right]_{aj2}^{*} \left[ V_{\chi Dd}^R \right]_{ai2}^{*} -  (Z_D)_{5i}^{*}(Z_D)_{2j}^{*} \left[ V_{\chi Dd}^L \right]_{aj3} \left[ V_{\chi Dd}^R \right]_{ai3} \right\}  \\
	- \frac{g_3^2}{16\pi^2} \frac{\mneu{a} \mg}{3} D_0(\mD{i}^2,\mD{j}^2,\mneu{a}^2,\mg)  \\
	\times \left\{ (Z_D)_{3j}(Z_D)_{5i}^{*} \left[ V_{\chi Dd}^R \right]_{ai3} \left[
V_{\chi Dd}^L \right]_{aj2}^{*} + (Z_D)_{6j}(Z_D)_{2i}^{*} \left[ V_{\chi Dd}^L
\right]_{ai3} \left[ V_{\chi Dd}^R \right]_{aj2}^{*} \right\} ~.\nonumber
\end{multline}
To obtain the coefficients $\tilde C_{1,2,3}^{\tilde g \chi 0}$ from 
$C_{1,2,3}^{\tilde g \chi 0}$, one again has to interchange left- with right-handed 
neutralino couplings. In addition, analogous replacements have to be performed for 
the elements of the down squark mixing matrix that explicitly appear in the above 
expressions: $(Z_D)_{2i} \leftrightarrow (Z_D)_{5i}$ and $(Z_D)_{3i} \leftrightarrow (Z_D)_{6i}$.

\subsubsection*{Gluino contributions}
%
\begin{multline}
	C_1^{\tilde g \tilde g} = -\frac{g_3^4}{16\pi^2} \frac{1}{9} \mg^2 D_0(\mD{i}^2,\mD{j}^2,\mg^2,\mg^2) (Z_D)_{3i}(Z_D)_{3j}(Z_D)^{*}_{2i}(Z_D)^{*}_{2j}  \\
	- \frac{g_3^4}{16\pi^2} \frac{11}{9} D_2(\mD{i}^2,\mD{j}^2,\mg^2,\mg^2) (Z_D)_{3i}(Z_D)_{3j} (Z_D)^{*}_{2i}(Z_D)^{*}_{2j}~, \nonumber
\end{multline}
%
%
\begin{equation}
	C_2^{\tilde g \tilde g} = - \frac{g_3^4}{16\pi^2} \frac{17}{18} \mg^2 D_0(\mD{i}^2,\mD{j}^2,\mg^2,\mg^2)(Z_D)_{3i}(Z_D)_{3j}(Z_D)^{*}_{5i}(Z_D)^{*}_{5j}~,\nonumber
\end{equation}
%
%
\begin{equation}
	C_3^{\tilde g \tilde g} = \frac{g_3^4}{16\pi^2} \frac{1}{6} \mg^2 D_0(\mD{i}^2,\mD{j}^2,\mg^2,\mg^2) (Z_D)_{3i}(Z_D)_{3j}(Z_D)^{*}_{5i}(Z_D)^{*}_{5j}~,\nonumber
\end{equation}
%
%
\begin{multline}
	C_4^{\tilde g \tilde g} = - \frac{g_3^4}{16\pi^2} \frac{7}{3} \mg^2 D_0(\mD{i}^2,\mD{j}^2,\mg^2,\mg^2) (Z_D)_{3i}(Z_D)_{6j}(Z_D)^{*}_{2i}(Z_D)^{*}_{5j}  \\
	+ \frac{g_3^4}{16\pi^2} \frac{2}{9} D_2(\mD{i}^2,\mD{j}^2,\mg^2,\mg^2) (Z_D)_{3i}(Z_D)_{6j} \left\{ 6 (Z_D)^{*}_{2i}(Z_D)^{*}_{5j} + 11 (Z_D)^{*}_{2j}(Z_D)^{*}_{5i} \right\}~,\nonumber
\end{multline}
\begin{multline}
	C_5^{\tilde g \tilde g} = -\frac{g_3^4}{16\pi^2} \frac{1}{9} \mg^2 D_0(\mD{i}^2,\mD{j}^2,\mg^2,\mg^2) (Z_D)_{3i}(Z_D)_{6j}(Z_D)^{*}_{2i}(Z_D)^{*}_{5j}  \\
	+ \frac{g_3^4}{16\pi^2} \frac{10}{9} D_2(\mD{i}^2,\mD{j}^2,\mg^2,\mg^2) (Z_D)_{3i}(Z_D)_{6j} \left\{ 3 (Z_D)^{*}_{2j}(Z_D)^{*}_{5i} - 2 (Z_D)^{*}_{2i}(Z_D)^{*}_{5j} \right\}~.\nonumber
\end{multline}
To get the corresponding expressions for the coefficients $\tilde C_{1,2,3}^{\tilde g \tilde g}$ one 
again has to carry out the following replacements for the down squark mixing matrix: 
$(Z_D)_{2i} \leftrightarrow (Z_D)_{5i}$ and $(Z_D)_{3i} \leftrightarrow (Z_D)_{6i}$.


\subsection{Chargino and Neutralino couplings} \label{app:XN-couplings}
Here we explicitly report -- in terms of gauge couplings, Yukawa couplings and rotation
matrices -- the chargino and neutralino couplings used in the expressions for the Wilson 
coefficients. 
\begin{equation}
	\left[V_{\chi Ud}^L \right]_{aiI} =  \left( - \frac{e}{s_W } (Z_{U})_{Ki}^{*}
(Z_{+})_{1a} + \hat Y_u^K (Z_{U})_{(K+3)i}^{*} (Z_{+})_{2a} \right) K_{KI}~,\nonumber
\end{equation}
\begin{equation}
	 \left[V_{\chi Ud}^R \right]_{aiI} =  - \hat Y_d^I (Z_{U})_{Ki}^{*} (Z_{-})_{2a}^{*} K_{KI}~,\nonumber
\end{equation}
\begin{equation}
	 \left[V_{\chi Dd}^L \right]_{aiI} =  - \frac{e}{\sqrt{2} s_W c_W} (Z_{D})_{Ii}
\left( \frac{s_W}{3} (Z_{N})_{1a} - c_W (Z_{N})_{2a} \right) + \hat Y_d^I (Z_{D})_{(I+3)i} (Z_{N})_{3a}~,\nonumber
\end{equation}
\begin{equation}
	 \left[V_{\chi Dd}^R \right]_{aiI} =  - \frac{ e\sqrt{2} }{3 c_W} (Z_{D})_{(I+3)i}
(Z_{N})_{1a}^{*} + \hat Y_d^I (Z_{D})_{Ii} (Z_{N})_{3a}^{*}~.\nonumber
\end{equation}
%
\subsection{Loop functions}
Finally, we give the explicit expressions for the loop functions that appear in the Wilson
coefficients listed above.
\begin{multline}
 	D_0(m_1^2,m_2^2,m_3^2,m_4^2) = \frac{m_1^2 \ln{m_1^2}}{(m_4^2-m_1^2)(m_3^2-m_1^2)(m_2^2-m_1^2)}  \\
	+\{1 \leftrightarrow 2\} + \{1 \leftrightarrow 3\} + \{1 \leftrightarrow 4\}~,\nonumber
\end{multline}
\begin{multline}
 	D_2(m_1^2,m_2^2,m_3^2,m_4^2) = \frac{1}{4} \left[ \frac{m_1^4 \ln{m_1^2}}{(m_4^2-m_1^2)(m_3^2-m_1^2)(m_2^2-m_1^2)} \right. \\
	+\{1 \leftrightarrow 2\} + \{1 \leftrightarrow 3\} + \{1 \leftrightarrow 4\} \bigg]~.\nonumber
\end{multline}


\bibliography{biblio} 
\bibliographystyle{JHEP}

\end{document}